**Rheological Behavior of Colloidal Silica Dispersion: Irreversible Aging and Thixotropy**

Vivek Kumar, Yogesh M. Joshi*

Department of Chemical Engineering, Indian Institute of Technology Kanpur, Kanpur, 208016, India

*Corresponding author

Email: joshi@iitk.ac.in

**Abstract**

In this work, we study the rheological behavior of colloidal dispersion of charge-screened nanoparticles of silica suspended in aqueous media that exhibits soft solid-like consistency. We observe that the system shows various characteristics of physical aging wherein it undergoes time evolution of rheological properties such as elastic modulus, relaxation time, and yield stress subsequent to shear melting of the same. Notably, the relaxation time increases more strongly than linearly with time, which is suggestive of hyper-aging dynamics. When considered along with the time-dependent yield stress, this behavior indicates the steady state shear stress-shear rate flow curve to be non-monotonic with a negative slope in a lower shear rate region. Performing shear melting on this system at a later date since the preparation of the dispersion (rest time) results in higher viscosity as well as yield stress, and the corresponding evolution of the elastic modulus shifts to lower times. This implies that physical aging in studied silica dispersion, while reversible over short time scales (of the order of hours), becomes irreversible over longer durations (days) owing to the inability of strong shear to break interparticle bonds that have strengthened over long durations. We also develop a thixotropic structural kinetic model within a time-dependent Maxwell framework that captures the experimentally observed rheological behavior well.

**Introduction**

Colloidal gels are an important class of soft materials formed by the aggregation of particulate matter suspended in a fluid, creating a space-spanning network that exhibits unique physical properties.[1] Colloidal gels have a significant breadth of industrial applications, ranging from food science,[2] pharmaceuticals,[3] agrochemicals,[4] home and personal care materials,[5] construction materials,[6, 7] energy storage,[8] etc. Their versatility originates from the wide range of particle sizes and shapes, as well as the tunability of inter-particle interactions and suspending fluid characteristics, which together give rise to diverse microstructures and rheological behaviors.[1, 9-15] Interestingly, although the suspended particles are thermal in nature, their mobility in the gel state gets significantly restricted, which renders them limited access to the phase space.[9, 16, 17] Consequently, the system falls out of thermodynamic equilibrium, initiating very slow microstructural rearrangements (continuation of network formation and strengthening) to progressively lower the free energy of the same.[9, 17, 18] This process is termed as physical aging, wherein the rheological properties of the gel evolve with time, adding another dimension to their physical behavior. For some colloidal gels, the process of physical aging can be reversed by applying a sufficiently strong deformation field, leading to shear rejuvenation or melting through network breakdown. This dynamic behavior, where the network forms under quiescent or weak flow (physical aging), leading to an increase in viscosity, and breaks down under strong flow (shear rejuvenation), resulting in a decrease in viscosity, is termed "*thixotropy*" in the rheology literature.[19, 20] Consequently, beyond system-specific characteristics, the rheological properties of colloidal gels are governed by the interplay between network formation and its strengthening (or physical aging) during rest or weak flow conditions, and the destruction of that network under more intense shear conditions. This competition leads to a variety of complex rheological phenomena widely reported in the literature.[21-34] Among the diverse classes of colloidal gels, those based on silica nanoparticles hold particular significance due to their easy availability, exceptional stability, and broad applicability. In this article, we investigate the interplay between their microstructural evolution, destruction, and how the storage time influences their rheological properties, offering insights into optimizing their design for advanced applications.

Silica, also known as silicon dioxide, has a chemical formula SiO$_2$. In an aqueous environment, the surface of the electrostatically stabilized silica possesses the silanol group (≡Si-OH).[6, 35-37] At high pH (well above the point of zero charge, which is typically below the pH of 2 for silica), the silanol group deprotonates and silica nanoparticles carry a negative surface charge.[38] This negative surface charge creates the required electrostatic repulsion between the silica particles, resulting in the colloidal stability of the sol. Adding an electrolytic solution (such as aqueous NaCl) to the dispersion starts the ion exchange between the electrolyte and the colloidal particles, exchanging the protons with the sodium ions.[6, 37, 39, 40] This process disrupts the hydrogen bonding between silica particles and water molecules and initiates the formation of inter-particle attractive interactions between the silica particles.[41-43] In addition, recent studies have shown that the surface heterogeneity on charged colloidal particles in the presence of salt may lead to the patchy charged domains, leading to weak-attractive interactions even between like-charged particles.[44] These attractions drive the formation of aggregates, which eventually form the percolated gel network. Furthermore, the close proximity between the silica particles in the gel network allows for siloxane bond formation between the interacting silica surfaces, leading to irreversible aggregation and the formation of larger non-Brownian aggregates.[41-43]

Many groups have carried out rheological studies on colloidal silica dispersion with particles having different diameters, volume fractions ($\phi$), and the types (monovalent, divalent, etc.) and concentrations of salts that induce gelation. Such salts have also been termed as accelerators. A detailed list of prior studies on colloidal silica dispersion, along with key observations, is given in Table 1. Cao et al.[45] studied the effect of varying $\phi$ and NaCl concentration on the gelation kinetics of 7 nm colloidal Ludox SM dispersion and observed that the gelation time (characterized by $G'$ and $G''$ cross-over) decreased with increasing $\phi$ or salt concentration. Interestingly, they also observed that the increased salt concentration only affected the kinetics, not the gel microstructure, since all the dynamic moduli curves merged at larger times elapsed since the cessation of shear melting (termed as $t_w$). Metin et al.[46] studied the gelation kinetics of the 5 nm colloidal 3 M silica dispersion with NaCl as the accelerator by varying the concentrations and observed that while in the concentration range of the silica ($\phi$ < 0.775) and NaCl (0.06 – 0.72 M), the system formed a stable gel, with further increase in the concentration of NaCl, the system transitioned into the two-phases, wherein solid gel phase at the bottom gets separated from a clear supernatant liquid at the top. They define the gel time at which $G'$ shows a sharp increase, which also corresponds to the time at which the UV-vis absorbance value attains a plateau. The subsequent increase in the $\phi$ at higher concentrations of NaCl (> 0.72 M) further affected the network strength, and the system ultimately transitioned into the viscous liquid state. Hatami et al.[47] used different accelerators (HCl, NaCl, KCl, $CaCl_2$, and $MgCl_2$) in the 0.002 – 0.22 M concentration range to study their effect on the gelation kinetics of Ludox SM colloidal silica. They observed that, within the explored concentration range, the increase in the concentration of accelerators or silica decreases the gelation time (defined by $G'$ and $G''$ cross-over). Interestingly, with increasing pH, initially, the gelation time decreased, but a further increase in the pH subsequently increased the gelation time, resulting in a minimum. Interestingly, the authors also observed that the addition of NaCl and HCl in combination with $CaCl_2$ or $MgCl_2$ anomalously decreased the rate of gelation. They argue that increased ionic strength of the solution due to the addition of NaCl or HCl decreases the chemical potential of calcium or magnesium ions, resulting in delayed gelation. Interestingly, some clay dispersions also show a decrease in gel time with an increase in concentration of multivalent ions.[48]

**Table 1: List of rheological studies on aqueous colloidal silica dispersions**

| Research Group | Silica systems, Accelerators, and the volume fraction of colloidal silica | Characterization Technique | Key Observations |
|---|---|---|---|
| **Manley et al., 2005**[49] | 10 nm Colloidal silica<br>MgCl$_2$<br>$\phi$ = 0.02 - 5 % | Rheometer, SLS, and DLS | $G'$ showed power law dependence on waiting time ($G' \sim t_w^{0.4}$) and $\phi$ ($G' \sim \phi^{3.6}$). |
| **Okazaki et al., 2008**[50] | 13 nm Snowtex 20 silica<br>LiCl, NaCl, and KCl<br>$\phi$ = 0.5 – 5.1 % | Rheometer | At fixed salt concentration, $G'$ increased with $\phi$. For fixed $\phi$, $G'$ initially increased with the salt concentration, but at higher salt concentrations, moduli decreased, and finally, at even higher concentrations (for $C_{\text{NaCl}}$ >1.5 M and $C_{\text{KCl}}$ >0.8 M), the system separated into two different phases. Additionally, the yield strain showed a decreasing power law dependence on $\phi$. |
| **Møller et al., 2008**[51] | 20 nm Ludox TM-40 silica<br>NaCl<br>$\phi$ = 6 % | Rheometer coupled with Magnetic resonance image velocimeter | The cone and plate geometry was used to apply a homogeneous stress field, showing that shear banding in aging thixotropic systems results from "age" heterogeneity, not stress heterogeneity. |

| **Cao et al., 2010**[45] | 7 nm Ludox SM silica<br>NaCl<br>$\phi$ = 1 – 4.7 % | Rheometer, SLS, and DLS | Gelation time decreased with an increase in the salt concentration or $\phi$. |
|---|---|---|---|
| **Metin et al., 2014**[46] | 5 nm Colloidal 3 M silica<br>NaCl<br>$\phi$ = 0.5 – 1.95 % | Rheometer | The system formed a stable gel for $\phi <$ 0.775 and NaCl concentration of 0.06 – 0.72 M. At higher NaCl concentrations (> 0.72 M), it transitioned into two phases. For NaCl concentration (> 0.72 M) and $\phi >$ 0.775, it became a viscous liquid. |
| **Kurokawa et al., 2015**[26] | 20 nm Ludox TM-40 silica<br>NaCl<br>$\phi$ = 6 % | Rheometer combined with ultrasonic velocimeter | Below the critical shear rate, the system exhibited steady-state shear banding. Above it, transient shear banding occurred initially, followed by homogeneous flow. The critical shear rate showed a non-linear dependence on waiting time, remaining constant initially, followed by an increase. |
| **Hatami et al., 2021**[47] | 7 nm Ludox SM silica<br>HCl, NaCl, KCl, $CaCl_2$, $MgCl_2$<br>$\phi$ = 2.07 – 3.7 % | Rheometer | The gelation time decreased with increasing concentrations of silica, NaCl, KCl, $CaCl_2$, and $MgCl_2$. It showed non-monotonic behavior with pH, with a minimum at the optimum pH. Adding NaCl and HCl with $CaCl_2$ or $MgCl_2$ decreased the gelation rate. |
| **Sögaard et al., 2022**[36] | Colloidal silica<br>Levasil CS40-213<br>Levasil CS40-222 | Rheometer | Smaller silica particles gelled faster than the larger ones due to lower activation energy requirements. Gel |
| | Levasil CS30-236 NaCl $\phi$ = 14.63 % | | time also decreased with increased temperature. |
| **Ghaffari et al., 2022**[52] | 10 – 30 nm Colloidal silica NaCl $\phi$ = 0.38 – 2.35 % | Rheometer | Depending on NaCl and silica concentration, the dispersion existed in the following phases: clear suspension, single-phase gel, two-phase gel, turbid suspension, and viscous suspension. Stronger single-phase gels had a lower yield strain (fragile) than the weaker single-phase gels. |
| **Kumar et al., 2025** | 20 nm Ludox TM-40 silica NaCl $\phi$ = 6 % | Rheometer | The silica gel system exhibits hyper-aging dynamics, with relaxation time increasing faster than linearly with time. Due to its thixotropic aging nature, the yield stress also increases with time. Consequently, the system shows non-monotonic flow curve behavior, with a critical shear rate separating the homogeneous and shear-banded parts of the flow curve. Over long rest (or storage) times, due to the formation of strong interparticle bonds, the system shows irreversible aging dynamics, leading to increased viscosity, slower evolution of relaxation time, higher critical shear rate, and enhanced yield stress. |

Sögaard et al.[36] explored the effect of the diameter of the colloidal silica particles (Levasil CS40 – 213, Levasil CS40 – 222, and Levasil CS40 – 236) on gelation kinetics using NaCl as an accelerator and observed that the decrease in the particle diameter decreases the gelation time (the gel time was characterized by tilting the sample by 90°, which is corroborated by a sudden increase in the complex viscosity). Interestingly, they also observed that the rise in temperature decreases the gelation time. Recently, Ghaffari et al.[52] studied the effect of $\phi$ and NaCl concentration on gel formation and observed that depending on the concentration of the silica (particle diameter = 10 – 30 nm) and NaCl, the system exists in the following phases: clear suspension, single-phase gel, two-phase system, turbid suspension, and viscous suspension. Interestingly, the authors observed that the stronger single-phase gels had a lower yield strain (more fragile) than the weaker single-phase gels.

Manley et al.[49] studied the effect of $t_w$ and $\phi$ on the evolution of $G'$ for the gel prepared using $MgCl_2$ as an accelerator for the 10 nm colloidal silica particles at both the Earth and the International Space Station (ISS). They observed that irrespective of the $\phi$, $G'$ showed a constant power law dependence on $t_w$ ($G' \sim t_w^{0.4}$), similarly at constant $t_w$, $G'$ showed a constant power law dependence on $\phi$ ($G' \sim \phi^{3.6}$), based on these observations, the authors concluded that the increase in the strength of the local bond elasticity between the silica particles causes an increase in the $G'$. Okazaki et al.[50], explored the effect of $\phi$ for the 13 nm Snowtex 20 colloidal silica in combination with different types of accelerators (LiCl, NaCl, and KCl) and their concentrations. They observed that above the critical coagulation concentration of the accelerator, the $G'$ increased with increasing $\phi$. Interestingly, for the fixed $\phi$, while the $G'$ initially increased with increasing concentration of the accelerators, eventually, above a critical concentration of the accelerator, it started decreasing, and finally, the system separated into two different phases. Additionally, the yield strain also showed a decreasing power law dependence on the $\phi$.

Interestingly, silica dispersion shows various interesting rheological behaviors in addition to the classic sol-gel transition, particularly with respect to its soft glassy rheology behavior. Møller et al.[51] using magnetic resonance image velocimetry coupled with the rheometer explored the effect of age heterogeneity on the formation of shear bands in the system using the cone and plate geometry (homogeneous stress field) and concluded that in the aging system, shear banding is observed due to the aging heterogeneity and not due to the stress heterogeneity. More importantly, they also showed that the system exhibits a non-monotonic flow curve below a critical shear rate, wherein the stress decreases with increasing shear rate. Subsequently, Kurokawa et al.[26] explored the shear banding behavior in finer detail using ultrasonic velocimetry combined with the rheometer and observed that while for shear rates applied below a critical shear rate the system showed the presence of the permanent shear band, for shear rates applied above the critical shear rate the system showed the presence of transient shear banding for some time at the start of the experiment and subsequently flowed homogeneously. While exploring the effect of waiting time ($t_\text{w}$) on the critical shear rate, they observed that the critical shear rate shows a non-linear dependence on $t_\text{w}$, wherein the critical shear rate remained constant at smaller $t_\text{w}$ and subsequently started to increase with an increase in the $t_\text{w}$.

In the present manuscript, we systematically investigate the physical aging and thixotropic behavior of colloidal silica dispersions formed via salt-induced gelation of Ludox TM-40 nanoparticles. While physical aging and thixotropy are well-documented in colloidal gels, the extent to which aging becomes irreversible over extended rest periods, and the implications of such irreversibility on rheological behavior, remain insufficiently understood. We examine how rest time influences the evolution of viscosity, elastic modulus, relaxation time, and yield stress. We also study whether shear rejuvenation remains effective after prolonged quiescence. We further explore the time-dependent flow behavior of a system, including the development of non-monotonic stress–shear rate curves, and how the physical aging history of the material affects the same. To interpret the experimental findings and provide a mechanistic understanding of the observed dynamics, we use a structural kinetic model within a time-dependent Maxwell framework. This modeling approach supports the analysis of competing processes of microstructural buildup and breakdown, offering a unified framework to describe the aging and flow behavior of charge-screened colloidal silica gels.

**Material and Methods**

Sample Preparation: In this work, Ludox® TM-40 colloidal silica (Sigma Aldrich) is used to prepare the gel. The sample preparation procedure was adapted from Kurokawa et al..[26] We used ultrapure water (resistivity 18.2 MΩ.cm) to prepare a 10 wt.% aqueous NaCl solution, which was mixed with a stock silica suspension in the 13:6 weight ratio using the ULTRA TURRAX T-25 homogenizer for 2 minutes at 9500 rpm. The resulting colloidal dispersion has a silica volume fraction ($\phi$) of 6 vol.% and an ionic strength of approximately 1.4 M. Interestingly, initially the mixture remains in the sol state, but as the mixing proceeds, within two minutes the system transitions into the gel state. The sample was then stored under quiescent conditions in airtight sealed polypropylene bottles. Rheological experiments were conducted on samples at different time intervals after sample preparation, referred to as the rest time ($t_R$ = 1, 6, 12, 18, and 24 days).

Rheological measurements: Rheological measurements were performed on an Anton Paar MCR-501 rheometer. The serrated Couette cell with a bob (diameter: 26.650 mm, length: 40.020 mm) and cup with a 1.1325 mm gap were used for all the experiments. To ensure an identical state before the experiments, all samples were shear melted at a high shear rate of 500 s$^{-1}$ for 1000 s to erase the prior deformation history. Subsequently, the following experiments were performed on samples stored for different rest times (1 to 24 days) to assess the rheological behavior of the silica gel system.

- Cyclic frequency sweep experiments: These experiments probe the evolution of dynamic moduli ($G'$ and $G''$) with time over the range of angular frequencies ($\omega \in [1, 46.42]$ rad/s) in the linear domain at 0.1% strain amplitude for 10000 seconds. For all the systems stored for different rest times, these experiments were done at 25 °C. For 1-day-old silica gel, we also studied the effect of temperature, wherein we performed the same experiments from $T = 25$ to 50 °C at an interval of 5 °C.
- Creep experiments: To understand the evolution of relaxation time ($\tau(t)$) with time, independent creep and recovery experiments, respectively, at 0.5 Pa and 0 Pa were performed at different waiting times elapsed since shear melting ($t_w$) as well as different rest times ($t_R$),

- Strain sweep experiments: The static yield stress ($\sigma_y$) was obtained by performing strain sweep experiments at a constant angular frequency of 10 rad/s at different $t_w$.
- Viscosity bifurcation: Following the shear melting step, the system was aged for 10 s and subsequently subjected to constant shear stress (CSS) in the range of 0.5 Pa to 10 Pa for 1000 seconds, and the corresponding evolution of shear rate was measured.
- Constant shear rate: To generate the steady-state shear stress–shear rate flow curve, similar to the constant shear stress (CSS) deformation field, the constant shear rate (CSR) deformation field was also applied to the system for 1000 seconds.

During the waiting period ($t_w$), a small oscillatory deformation field with 0.1% strain amplitude and 0.63 rad/s angular frequency is applied to the system. To avoid evaporation, the free surface of the sample in the shear cell was covered with low-viscosity silicon oil. Unless otherwise mentioned, all the experiments were performed at 25 °C.

Zeta potential: For the zeta potential measurement, the stock Ludox® TM-40 colloidal silica with 40 wt.% silica was diluted 20 times in the Millipore water. The original pH of this diluted sample was 9.7. Therefore, HCl was added to the diluted solution to prepare samples of different pH of lower values. Subsequently, the zeta potential of the system was measured using the Malvern Nano ZS Zetasizer. Field Emission Scanning Electron Microscopy (FESEM): For the FESEM imaging, the stock Ludox® TM-40 colloidal silica was diluted 250 times in the Millipore water. The diluted sample was drop-cast onto the copper tape and left to dry. Subsequently, the FESEM image of the silica colloid was taken by SUPRA 40 VP, ZEISS. To determine the mean particle diameter, the acquired micrographs were analyzed using ImageJ software, enabling quantitative image-based particle size analysis.

**Results and Discussion**

We begin by characterising the colloidal silica dispersion studied in this work through Field Emission Scanning Electron Microscope (FESEM) imaging and Zeta potential measurements. Fig. 1 shows an FESEM image of Ludox TM-40 colloidal silica; it can be seen that the monodispersed colloidal silica particles have a mean diameter of 20 $\pm$ 1.1 nm (refer to supporting information Fig. S1 for the particle size distribution histogram). This observation and the image are consistent with those reported by Kurokawa et al..[26] In Fig. 2, we plot the zeta potential of Ludox TM-40 colloidal silica as a function of pH. At higher pH values, deprotonation leads to the formation of negatively charged $SiO^-$ groups on the silica surface, resulting in a negative zeta potential. However, upon decreasing the pH of the dispersion (through the addition of HCl), the relative increase in the concentration of the $H^+$ ions in the dispersion reduces the extent of deprotonation of silica particles (less negatively charged surface), leading to a reduction in the magnitude of the zeta potential. Notably, from the zeta potential measurement, we observe that the point of zero-charge of the Ludox TM-40 colloidal silica is less than pH $=$ 1, which matches well with that mentioned in the literature.[38] The zeta potential-pH profile in Fig. 2 characterizes the electrostatic behavior of Ludox TM-40 colloidal silica in the absence of added salt. At the initial pH of 9.7, the surface silanol groups ($\equiv Si - OH$) are extensively deprotonated, resulting in highly negative surface charges that ensure strong electrostatic repulsion between the colloidal silica particles and stabilization of the dispersion. Upon the addition of 1.4 M NaCl, due to the reprotonation of some of the $\equiv Si - O^-$ groups, the system's pH decreases to 7.9. At this high salt concentration, the high ionic strength compresses the electrical double layer, effectively screening electrostatic repulsions. Concurrently, ion exchange between $Na^+$ ions and surface protons reduces the surface charge density.

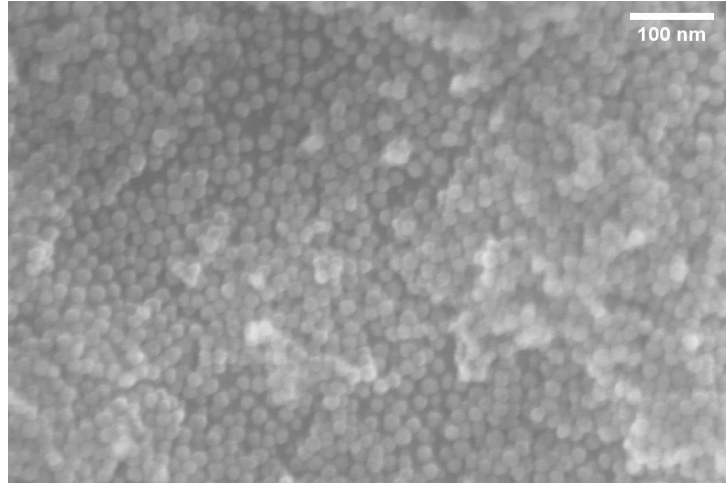

**Figure 1:** SEM image of Ludox TM-40 colloidal silica, which suggests the particles are in the 20±1.1 nm range.

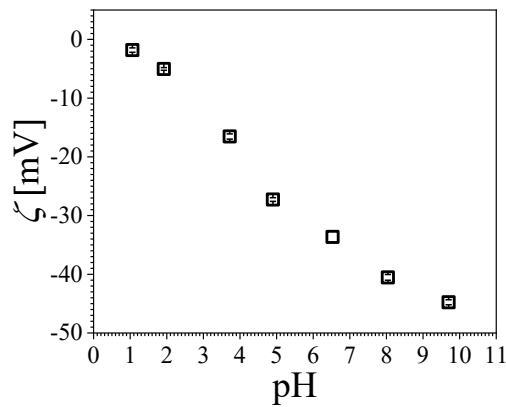

**Figure 2:** Variation in the zeta potential ($\zeta$) of Ludox TM-40 colloidal silica with pH. It can be seen that the point of zero charge of the silica particles is less than pH 1.

Soft glassy materials, owing to their thermodynamically out-of-equilibrium state, remember their past deformation history. Consequently, to get a reproducible initial state, a large deformation field (oscillatory or rotational) is usually applied before performing any rheological experiment.[52, 53] This process, known as shear melting or rejuvenation, destroys the existing structures and erases the system's memory. In the present study, we subject the sample to a shear rate of 500 s$^{-1}$ for 1000 s. Fig. 3 (a) shows that the viscosity decreases as a function of time until it reaches a steady state with a constant value ($\eta_s$). Such a decrease is suggestive of the breakdown of structure during rejuvenation. In Fig. 3 (b), we plot $\eta_s$ of the samples stored for different rest times ($t_R$). It can be seen that $\eta_s$ increases with an increase in $t_R$, suggesting that applying a large deformation field does not completely rejuvenate or shear melt the samples to the same state. This behavior implies an irreversible nature of the aging process in colloidal silica dispersion over a duration of days. It should be noted that samples did rejuvenate to the same state (same level of $\eta_s$) after repeated shear melting for the same $t_R$. This indicates that colloidal silica dispersion (Ludox gel) undergoes reversible aging over a short period but irreversible aging over long periods.

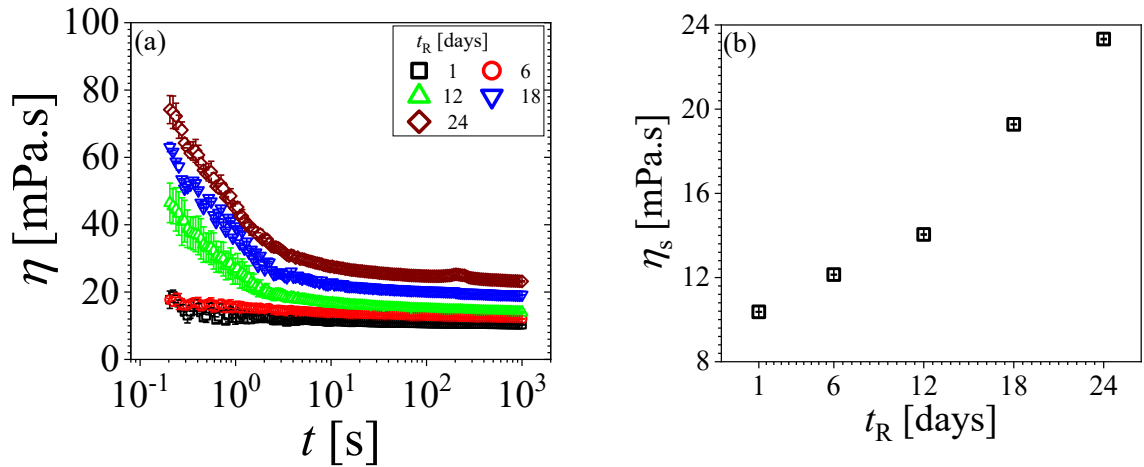

**Figure 3:** (a) Evolution of viscosity with time for the constant shear rate of 500 s$^{-1}$ is plotted for the system stored for different rest times ($t_R$) as mentioned in the legend. (b) The plateau value of viscosity ($\eta_s$) is plotted as a function of $t_R$.

We next conduct small amplitude oscillatory shear and creep experiments on samples stored for different rest times to investigate how the irreversible aging dynamics influence the system's microstructure and, consequently, the evolution of the system's dynamic moduli and relaxation time. In a small amplitude oscillatory or aging experiment, the system is subjected to the time-dependent oscillatory shear at different frequencies ($\omega \in [46.42 – 1]$ rad/s) following the cessation of shear melting to measure the evolution of elastic ($G'$), and viscous ($G''$) moduli with waiting time ($t_w$), defined as the time elapsed since shear melting stopped. Fig. 4 shows the evolution of the $G'$ and $G''$ as a function of $t_w$ for samples stored for different $t_R$ at an angular frequency ($\omega$) of 1 rad/s. It can be seen that, immediately after shear melting, $G'$ dominates over the $G''$ for all the $t_R$. Interestingly, for the fixed $t_R$, while $G'$ monotonically increases with $t_w$, and eventually shows a power law dependence on the same. The viscous modulus ($G''$), initially increases and then shows a weak prolonged decrease over the explored time duration. Interestingly, similar to the increasing viscosity ($\eta_s$) with $t_R$, $G'$ and $G''$ also increase and shift vertically to higher values with increasing $t_R$.

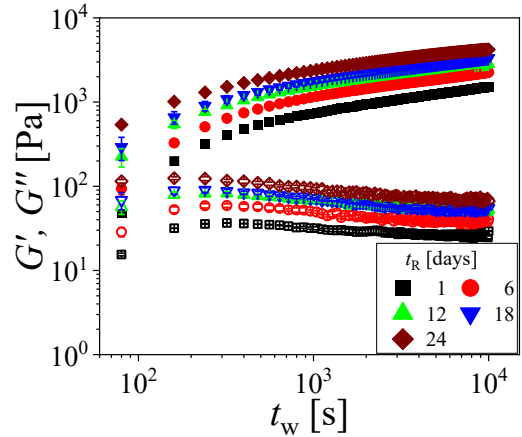

**Figure 4:** The evolution of elastic modulus ($G'$) (filled symbols) and viscous modulus ($G''$) (open symbols) with time ($t_w$) is plotted for systems stored for different $t_R$. The values of the rest time are given in the legend. The applied strain amplitude and angular frequency during the small-amplitude oscillatory shear experiment are 0.1% and 1 rad/s, respectively.

In Fig. 5, we plot the dynamic moduli $G'$ and $G''$ as a function of angular frequency ($\omega$) for samples aged for different $t_R$. To minimize the influence of aging on the measurement of the dynamic moduli during the frequency sweep test, the experiment was completed within 80 s, significantly shorter than the 3000 s waiting time for which the data is plotted. As observed in Fig. 5, $G'$ consistently remains higher than $G''$ at all $t_R$ and does not show any dependence on $\omega$. In contrast, $G''$ initially decreases with increasing $\omega$, reaches a minimum at 21.54 rad/s, and increases again. Interestingly, the inverse of the $\omega$ at which the $G''$ shows a minimum correlates with the constituent particles' rattling timescale within the cage formed by the neighboring particles,[55] which comes out to be around 0.046 s and remains constant with rest time.

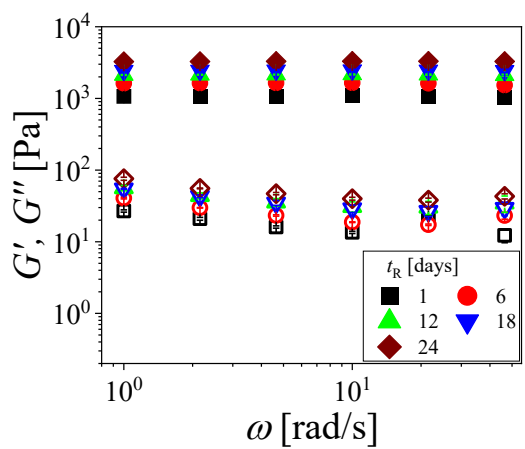

**Figure 5:** Elastic modulus ($G'$) (filled symbols) and viscous modulus ($G''$) (open symbols) are plotted as a function of the angular frequency ($\omega$) at $t_w = 3000$ s for different $t_R$. From bottom to top: $t_R = $ 1, 6, 12, 18, and 24 days. The applied strain amplitude during the frequency sweep experiment is 0.1%.

Subsequently, we also studied the effect of temperature on the gelation kinetics of the 1-day-old silica dispersion system. In Fig. 6 (a), we plot the time evolution of dynamic moduli at different temperatures. For temperatures ($T \leq 35$ °C), the dynamic moduli increase very weakly with temperature. Additionally, the elastic modulus shows a constant power law dependence on $t_w$ at larger $t_w$. However, interestingly, as the temperature increases further ($T \geq 40$ °C), after a temperature-dependent critical time, both $G'$ and $G''$ increase more intensely with time. The enhancement in both the dynamic moduli as a function of $T$ suggests expedited aging dynamics with an increase in $T$.

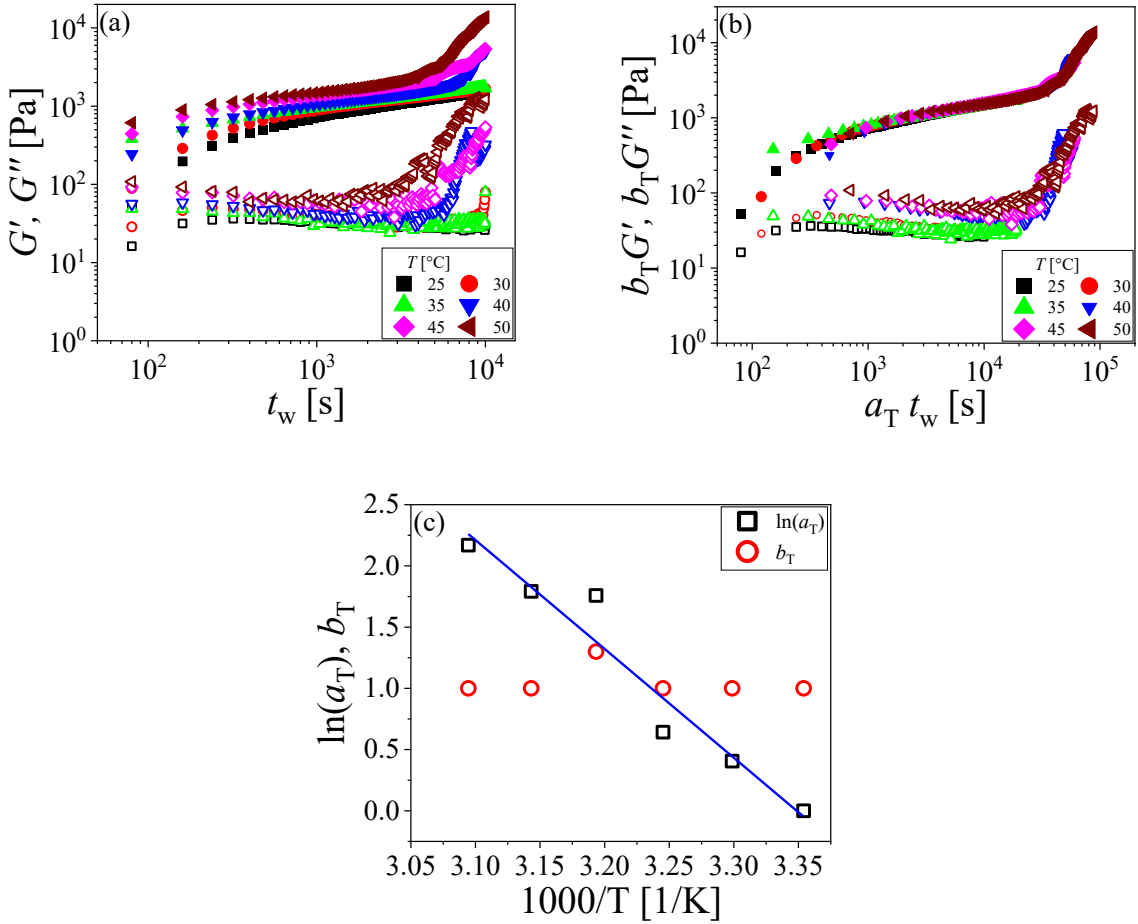

**Figure 6:** (a) The evolution of elastic modulus ($G'$) and viscous modulus ($G''$) is plotted as a function of $t_w$ for the 1-day-old sample at different temperatures. (b) Time – Temperature superposition of $G'$ and $G''$, after the vertical and horizontal shifting the dynamic moduli and $t_w$ data. (c) ln ($a_T$) and $b_T$ are plotted as a function of $1000/T$ with 25 °C as the reference temperature. The horizontal shift factor ($a_T$) shows the Arrhenius dependence on $T$. The applied strain amplitude and angular frequency during the oscillatory experiment are 0.1% and 1 rad/s, respectively. The values of different temperatures at which the aging experiments were conducted are given in the legend of (a) and (b).

It can be seen that the evolution of dynamic moduli plotted in Fig. 6(a) has self-similar curvature and shifts to lower times with an increase in $T$. In Fig. 6 (b), we plot horizontally and vertically shifted $G'$ and $G''$ data shown in Fig. 6 (a), leading to the time-temperature superposition of the $G'$ and $G''$ for the 1-day-old system. In Fig. 6 (c), we plot the corresponding shift factors $a_T$ (horizontal) and $b_T$ (vertical) as a function of $10^3/T$. It can be seen that $\ln a_T$ shows a linear dependence on $10^3/T$ while $b_T$ practically remains constant at unity. To analyze the temperature dependence of $a_T$, we need to understand how the evolution of modulus gets affected by the change in temperature. Typically, for soft glassy materials, the elastic modulus is observed to show the time dependence given by[56]:

$$G' = G_0 f(t_w/\tau_m), \qquad (1)$$

where $\tau_m$ represents the microscopic timescale, which refers to the characteristic timescale over which the internal structure of a soft glassy material undergoes rearrangement at the particle or molecular level. The microscopic timescale is proposed to demonstrate the Arrhenius relationship with temperature given by[56]:

$$\tau_m = \tau_{m0} \exp\left(\frac{U}{k_B T}\right), \qquad (2)$$

where $U$ denotes the energy barrier associated with the microscopic motion of the entity within the cage, $k_B$ is the Boltzmann constant, while $\tau_{m0}$ represents an attempt time. The generic dependence of elastic modulus on time and temperature given by Eq. (1) leads to: $a_T \propto 1/\tau_m$. Considering the relationship between $\tau_m$ and $T$ given by Eq. (2), we get $\ln a_T = \ln a_\infty - U/k_B T$. It can be seen that this expression is well expressed by a linear fit to the shift factor data as shown in Fig. 6(c), leading to $U \approx 75$ kJ/mol at the reference temperature 298.15 K. It should be noted that the typical value for $U$ for a covalent bond is around $250 - 750$ kJ/mol, while it is around 1 kJ/mol for the van der Waals interactions.[57] The estimated value of $U \approx 75$ kJ/mol suggests that the interaction energy between silica nanoparticles is stronger than the van der Waals interactions. It is important to clarify here that the extracted value of $U$ does not represent a single pairwise interaction between silica nanoparticles. Rather, it represents a mesoscopic energy barrier associated with the cooperative rearrangement of particles within the arrested gel network. In this system, the collective influence of multiple interactions, including van der Waals and covalent bonding, facilitates the formation of a percolated gel structure. Consequently, a single particle associates itself with multiple such interactions in order to have a percolated network which reflects in higher value of $U$.

The evolution of relaxation time is a ubiquitous feature of soft glassy materials. Creep or stress relaxation experiments are commonly used in soft glassy materials to quantify this time dependence.[58, 59] In this work, we employed the creep test protocol, wherein a constant stress ($\sigma$) is applied to colloidal silica dispersion at a certain rest time ($t_R$) after preparation at various waiting times ($t_w$) elapsed since shear melting. The corresponding evolution of creep compliance $\left(J(t - t_w)\right)$ is monitored as a function of creep time $(t - t_w)$. However, the induced compliance needs to be corrected for elastic strain accumulated during the shear rejuvenation.[60] To address this, the material is subjected to additional tests by applying zero stress at different waiting times ($t_w$) to obtain the recoverable strain. Following this procedure (explained in detail elsewhere),[60] the silica dispersion was subjected to 0.5 Pa and 0 Pa stress fields at each waiting time. The corrected creep compliance is then calculated by subtracting the recovered strain from the strain induced during the non-zero stress creep test. The corresponding corrected creep compliances were then calculated. Fig. 7 shows the corrected creep compliance curves for the 1-day-old samples as a function of creep time $(t - t_w)$ (for other $t_R$, refer to supporting information Fig. S2). It can be seen that lower compliance is induced in the material for an experiment started at higher waiting times. This behaviour has significant implications as far as its linear viscoelastic analysis is concerned. In typical systems that attain equilibrium, the time-translational invariance (TTI) principle holds in the real-time domain $(t - t_w)$.[1, 61] Therefore, in the creep experiments, the creep compliance curves $(J(t - t_w))$ remain independent of the waiting time $(t_w)$. Consequently, the creep curves obtained at different waiting times superimpose on each other without any vertical or horizontal shifting. However, in time-dependent and or thixotropic systems, both the relaxation time and modulus become time-dependent. In such systems, the TTI principle does not hold, and the creep compliances measured at different waiting times do not superimpose in the real-time domain. Consequently, $J(t - t_w)$ exhibits an additional dependence on $t_w$,[62] leading to $J = J(t - t_w, t_w)$ as shown in Fig. 7.

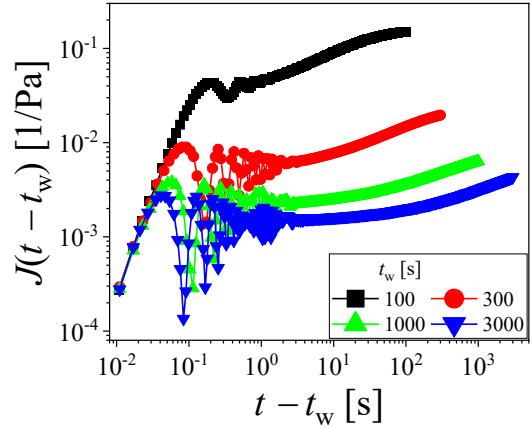

**Figure 7:** The corrected creep compliance $J(t - t_\mathrm{w})$ is plotted as a function of creep time $(t - t_\mathrm{w})$ for a 1-day-old colloidal silica dispersion at different $t_\mathrm{w}$.

For materials that follow the TTI principle in the real-time domain, the strain induced by a creep deformation field within the linear viscoelastic regime can be represented by the Boltzmann superposition principle, given by:[1, 61, 63]

$$\gamma(t - t_\mathrm{w}) = \int_{-\infty}^{t} J(t - t_\mathrm{w}) \frac{d\sigma}{dt_\mathrm{w}} dt_\mathrm{w} \tag{3}$$

where, $\gamma$ represents the strain generated in the material due to the application of the stress ($\sigma$) at $t = t_\mathrm{w}$. However, in time-dependent and/or thixotropic systems, the Boltzmann superposition principle fails to capture the material's behavior due to the failure of the TTI principle. To address this limitation, the concept of the effective time ($\xi(t)$) domain has been introduced,[64] wherein, the real-time is normalized by a time-dependent relaxation time ($\tau(t_\mathrm{w})$), such that the relaxation time of the system becomes constant in the effective time domain, allowing the system to follow the TTI principle within the effective time-domain framework. The effective time is expressed as:

$$\xi(t) = \tau_\mathrm{ETD} \int_{0}^{t} \frac{dt'}{\tau(t')} \tag{4}$$

where, $\tau_{\text{ETD}}$ is the constant relaxation time associated with the effective time domain, and $\tau(t')$ is the time-dependent relaxation time of the system. In the effective time domain, Boltzmann superposition principles can be rewritten as,[58, 64]

$$\gamma(\xi - \xi_{\text{w}}) = \int_{-\infty}^{\xi} J(\xi - \xi_{\text{w}}) \frac{d\sigma}{d\xi_{\text{w}}} d\xi_{\text{w}} \tag{5}$$

Joshi and co-workers [23, 56, 58-60, 65] have extensively used the concept of effective time domain for soft glassy materials to obtain the evolution of relaxation time on time.

In this method, in order to express the effective time domain, one needs to a priori assume a functional form of relaxation time dependence on time. There are various ways of obtaining this functional dependence a priori.[9, 23, 58, 65-70] However, for various glassy materials (including soft, amorphous polymeric, and spin glasses), it has been observed that the relaxation time typically exhibits a power law dependence on waiting time expressed as:[23, 58, 66-68]

$$\tau(t') = A\tau_{\text{m}}^{1-\mu} t'^{\mu} \tag{6}$$

where A is a constant, $\tau_{\text{m}}$ is a microscopic timescale related to a material, and $\mu$ is the power law exponent. Depending on the value of $\mu$, different soft glassy materials are defined as hyper-aging ($\mu > 1$), simple-aging ($\mu = 1$), and sub-aging ($\mu < 1$).[23, 60] After substituting the expression of $\tau(t')$ from Eq. (6) into Eq. (4) and then solving the integral for time $t$ and $t_{\text{w}}$, for $\tau_{\text{ETD}} = \tau_m$, we get:

$$\xi(t) - \xi(t_{\text{w}}) = \frac{\tau_m^{\mu}}{A} \left[ \frac{t^{1-\mu} - t_{\text{w}}^{1-\mu}}{1-\mu} \right]. \tag{7}$$

In Fig. 8 (a), we plot the vertically shifted corrected creep compliance ($G(t_{\text{w}})J(t - t_{\text{w}})$) as a function of $\left[ \frac{t^{1-\mu} - t_{\text{w}}^{1-\mu}}{1-\mu} \right]$ in the effective time domain (as shown) and it can be seen that the creep curves demonstrate a remarkable superposition for $\mu = 1.203 \pm 0.060$. The magnitude of the $\mu$ indicates that the 1-day-old sample exhibits hyper-aging behavior. The observation of superposition also corroborates that the Ludox colloidal silica system indeed follows the assumed power law functional dependence given by Eq. (6). Subsequently, following the same creep experimental protocol, we evaluate the effect of rest time on the relaxation behavior of the colloidal silica system. In Fig. 8 (b), the corresponding $\mu$ values are plotted as a function of rest time. It can be seen that $\mu$ decreases with increasing rest time as previously reported for Laponite® dispersion.[71] This behavior indicates that, although higher rest times result in increased relaxation times, the influence of waiting times on relaxation becomes progressively weaker.

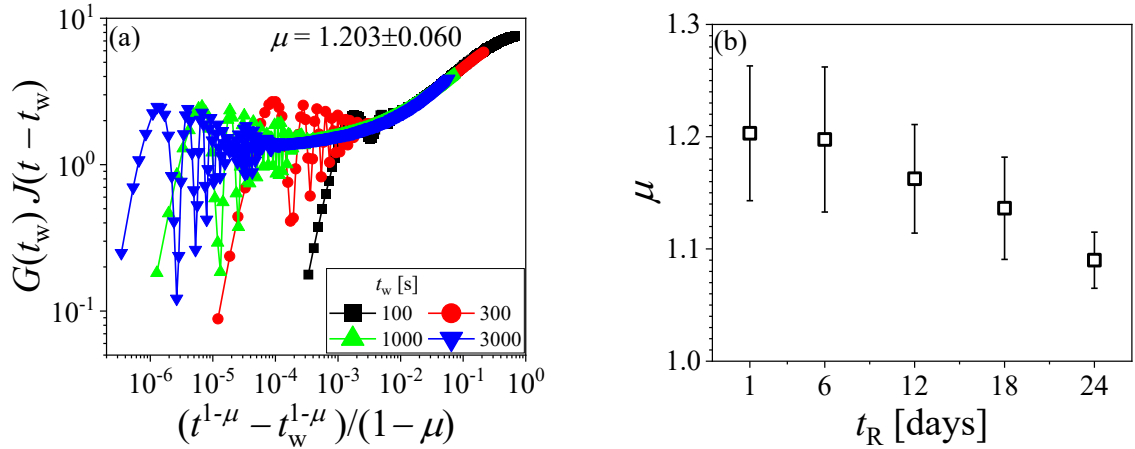

**Figure 8:** (a) Time – aging-time superposition of creep curves obtained in the effective time domain is plotted for $t_R = 1$. The values of different $t_w$ are given in the legend. (b) Power law coefficient parameter ($\mu$) necessary for obtaining the time aging time superposition is plotted as a function of the rest time ($t_R$).

We next explore how the rest time affects the steady-state flow curve of this hyper-aging system. To generate the steady-state flow curve of the silica dispersion, we subject the system to both constant shear rate (CSR) and constant shear stress (CSS) flow fields. Accordingly, in the creep protocol, the system is aged for 10 s, and subsequently, a constant stress is applied for 1000 s. In Fig. 9, the corresponding evolution of the shear rate at different stresses is plotted as a function of time for the 1-day-old sample (for other $t_R$, refer to supporting information Fig. S3). The system clearly shows two distinct behaviors depending on the magnitude of the applied stresses ($\sigma$). For $\sigma \leq 2$ Pa, the shear rate initially increases, achieves a maximum, and then starts decreasing, eventually leading to a stoppage of flow. However, for $\sigma \geq 3.05$ Pa, after the initial increase, the system finally attains a steady state and flows at a constant shear rate. This behavior has been termed as viscosity bifurcation in the literature, where the system ceases to flow below a critical stress (yield stress associated with viscosity bifurcation) but achieves a steady-state flow above it.[21, 23, 72-75] In this experiment, the yield stress associated with the viscosity bifurcation is 3.05 Pa, above which the system flows homogeneously. On similar lines, we subject colloidal silica dispersion on a certain rest day to a constant shear rate flow field and note the resultant steady-state stress.

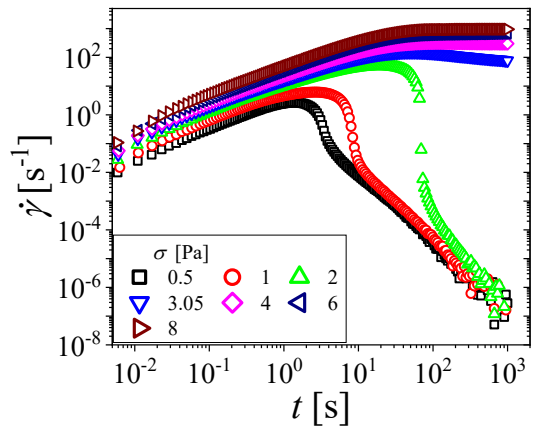

**Figure 9:** The evolution of shear rate ($\dot{\gamma}$) at different stresses is plotted as a function of time for a 1-day-old silica dispersion. The magnitude of different stresses is shown in the legend.

The fact that relaxation time and elastic modulus shows increase with respect to time implies viscosity of the studied silica dispersion increases under quiescent as well as weak stress conditions. The viscosity bifurcation experiments not just corroborates with that observation but also shows time dependent decrease in viscosity beyond a threshold stress. Both these observations clearly suggest that the present system shows all the characteristic features of thixotropy discussed in the literature.[19, 20, 23]

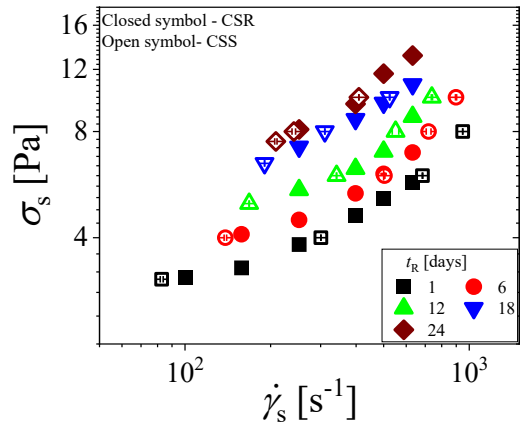

**Figure 10:** The steady-state shear stress is plotted as a function of steady-state shear rate (flow curves) for different $t_R$. The closed symbol is for the constant shear rate (CSR), whereas the open symbol is for the constant shear stress (CSS) experiments. The different $t_R$ values are given in the legend.

In Fig. 10, the experimental steady-state flow curve (steady-state shear stress versus shear rate) data is plotted for systems stored for different $t_R$. The closed symbols represent the data associated with the constant shear rate (CSR) flow field, while the open symbols depict the same related to constant shear stress (CSS) experiments. Since the viscosity bifurcation experiments suggest that below the yield stress associated with the same, no steady state shear rate can be maintained in a sample, this shear rate fixes the boundary of the homogeneous flow field. The shear rate and shear stress at the minima are designated as the critical values and are represented as $\dot{\gamma}_c$ and $\sigma_c$, respectively. In shear rate-controlled experiments, if the applied shear rate is below the $\dot{\gamma}_c$, indeed, a constant shear stress plateau is observed. But such shear rates ($\dot{\gamma} < \dot{\gamma}_c$) are not accessible in stress-controlled experiments and therefore may correspond to two possibilities: a steady state shear stress plateau in a limit of small shear rates (similar to that observed for Herschel-Bulkley fluid) or a non-monotonic flow curve wherein the flow curve shows a negative slope in a limit of small shear rates.[23, 26, 51] The latter is expected to form steady-state shear banding when the applied shear rate is less than the critical shear rate ($\dot{\gamma} < \dot{\gamma}_c$). It is practically impossible to fathom whether shear bands are present in a flow field unless one obtains the velocity field, which we do not in the present work. However, the major difference between the former (shear stress plateau in a limit of low shear rate) and the latter (non-monotonic flow curve) is the presence of time-dependent yield stress in the latter.[23, 76] In the case of the former, yield stress necessarily remains constant.[23, 76]

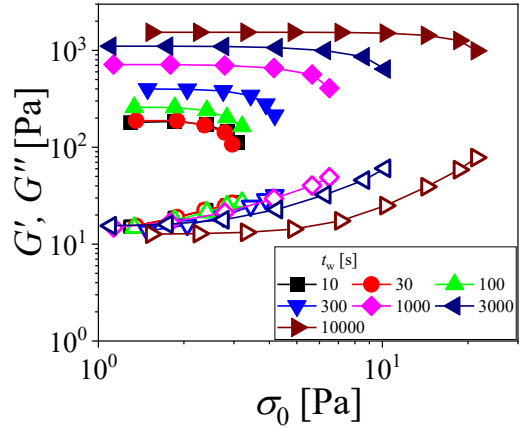

**Figure 11:** Elastic moduli ($G'$) (closed symbol) and viscous moduli ($G''$) (open symbol) are plotted as a function of stress amplitude ($\sigma_0$) at different waiting times ($t_w$) for 1-day-old silica dispersion. The angular frequency used during the amplitude sweep experiment is 10 rad/s. The different symbols represent the different waiting times (values are shown in the legend) at which the system is subjected to the amplitude sweep.

To determine the dependence of the yield stress on time ($t_w$) at different rest times $t_R$, colloidal silica dispersion is subjected to an oscillatory strain sweep flow field with a strain amplitude ranging from 0.1 to 10% at an angular frequency of 10 rad/s. In Fig. 11, corresponding $G'$ and $G''$ are plotted against the stress amplitude ($\sigma_0$) at different waiting times for the 1-day-old silica dispersion (for other $t_R$, refer to supporting information Fig. S4). Within the linear regime, at smaller stress amplitudes, both dynamic moduli remain constant. However, with increasing strain amplitude (and consequently, stress amplitude), the $G''$ starts increasing, whereas the $G'$ decreases. At a critical stress value, both the dynamic moduli and the stress induced in the system decrease sharply, indicating the yielding transition in the sample. The critical stress at which the system yields is referred to as the yield stress ($\sigma_y$) in the context of the amplitude sweep and is defined as the peak stress achieved by the sample prior to the sharp decline in both moduli at the yielding point. It can be seen in Fig. 11 that till the waiting time of $t_w \leq 100$ s, the change in $G'$ and $G''$ as a function of $\sigma_0$ overlaps but increases to higher values with an increase in the $t_w$.

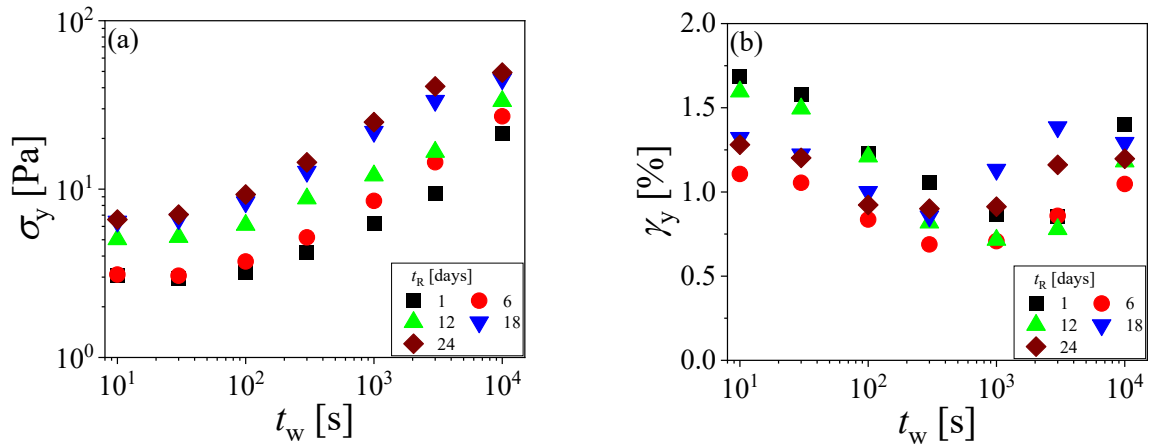

**Figure 12:** (a) Static yield stress ($\sigma_y$), and (b) yield strain ($\gamma_y$) is plotted as a function of waiting time ($t_w$) for the silica dispersion stored for different rest times ($t_R$).

In Fig. 12 (a), the experimentally measured static yield stress ($\sigma_y$) and yield strain ($\gamma_y$) values are plotted as a function of $t_w$ for different $t_R$. Notably, $\sigma_y$ remains almost constant for smaller $t_w$ (100 s for $t_R = 1, 6$ and 30 s for the remaining $t_R$) and subsequently starts increasing as $t_w$ increases. Furthermore, with an increase in $t_R$, the $\sigma_y$ curves shift vertically upward to the higher static yield stress value. The observed behavior, where the yield stress remains constant up to a certain waiting time before increasing, is a characteristic commonly observed in many soft glassy materials.[23, 77-80] Interestingly, the plateau value of the static yield stress obtained from the amplitude sweep and the yield stress obtained from the viscosity bifurcation ($t_w = 10$ s) show an excellent correspondence, where we observe similar values for yield stress measured through different methods. While the yield stress of the silica colloidal dispersion remains constant initially and then increases with the waiting time, the elastic modulus keeps on increasing with the waiting time, and consequently, as shown in Fig. 12 (b), we observe that the yield strain ($\gamma_y = \sigma_y/G'$) initially decreases and then starts increasing with the waiting time. The observation of time-dependent yield stress and strain further reinforces the present system to be thixotropic in nature.

The rheological behavior of colloidal silica dispersion, as discussed above, can be categorized into two distinct parts. The first part covers the various physical aging characteristics exhibited by the colloidal silica dispersion system, such as the time-dependent evolution of dynamic moduli, hyper-aging dynamics of the relaxation time, viscosity bifurcation, steady-state shear stress-shear rate flow curves with potential non-monotonicity, and time-dependent yield stress and strain. The second part focuses on how these behaviors evolve as a function of the rest day, which refers to the day after the preparation of the colloidal silica dispersion when the experiments were conducted. This latter aspect is associated with the irreversible aging dynamics exhibited by the system. Overall, the present system shows various characteristic features of physically aging thixotropic materials that can be modeled using a structural kinetic model, which are discussed in the present section.

The structural kinetic or $\lambda$ models are widely recognized for their ability to successfully capture the time-dependent and thixotropic behavior of soft glassy materials.[21, 23, 51, 81, 82] In these models, the parameter $\lambda$ serves as an abstract measure of the structural state of a material. Generally, $\lambda$ models consist of three key components: (i) Time evolution equation of the structure parameter ($\lambda$), representing the instantaneous structural state of the material under the application of the flow field. (ii) Constitutive equation describing the relationship between stress and strain rate. (iii) An expression relating the structure parameter ($\lambda$) to the rheological properties of the material, such as viscosity, relaxation time, elastic modulus, etc. In this work, we employ the structure kinetic model initially introduced by Coussot et al.[21, 51] and modified by Joshi[23] to introduce viscoelasticity, wherein the evolution equation for $\lambda$ is given by,

$$\frac{d\lambda}{d\bar{t}} = \frac{1}{T_0} - \frac{\beta\lambda}{\bar{\eta}(\lambda)}\bar{\sigma} \qquad (8)$$

The right side of the above equation contains two terms. The first term, $1/T_0$, represents the microstructural build-up (aging) term with $T_0$ as the thixotropic time scale associated with the system.[23] The smaller the value of $T_0$; the faster the aging. The second term is the microstructural breakdown term, where $\bar{\sigma}/\bar{\eta}(\lambda)$ is the viscous part $(\bar{\dot{\gamma}}_v)$ of the total shear rate $(\bar{\dot{\gamma}})$ and $\beta$ is a scalar parameter. In this model, $\lambda = 0$ is the structureless state, which is associated with the completely rejuvenated or shear-melted state of the material in the limit of the strong deformation field. As structure builds up, $\lambda$ increases indefinitely and eventually approaches $\infty$ in a limit of fully structured or the thermodynamic equilibrium state.[82] Furthermore, in Eq. 8, the microstructural destruction term has been made proportional to the viscous part $(\bar{\dot{\gamma}}_v)$ of the total shear rate $(\bar{\dot{\gamma}})$ and not the total shear rate $(\bar{\dot{\gamma}})$ to avoid violation of the second law of thermodynamics as discussed elsewhere.[31]

To model the rheological behavior, we use the time-dependent Maxwell model as the constitutive equation, which relates the shear stress $(\bar{\sigma})$ to the shear rate $(\bar{\dot{\gamma}})$ given by:

$$\frac{d}{d\bar{t}}\left(\frac{\bar{\tau}(\lambda)}{\bar{\eta}(\lambda)}\bar{\sigma}\right) + \frac{\bar{\sigma}}{\bar{\eta}(\lambda)} = \bar{\dot{\gamma}}_e + \bar{\dot{\gamma}}_v = \bar{\dot{\gamma}} \qquad (9)$$

where $\bar{\tau}(\lambda)$ and $\bar{\eta}(\lambda)$ are respectively the relaxation time and viscosity. The differential term on the right-hand side of Eq. (9) represents the instantaneous elastic shear rate $(\bar{\dot{\gamma}}_e)$, whereas the second term represents the instantaneous viscous shear rate $(\bar{\dot{\gamma}}_v)$. Finally, the parameters of the constitutive equation $\bar{\eta}$ and $\bar{\tau}$ can be related to the parameter $\lambda$ through:

$$\bar{\eta}(\lambda) = \eta_0(1 + \lambda^n), and \qquad (10)$$
$$\bar{\tau}(\lambda) = \tau_0(1 + \lambda^m) \qquad (11)$$

where $\eta_0$ and $\tau_0$ are, respectively, the constant viscosity and relaxation time of the system in the structure-less state ($\lambda = 0$), and $n$ and $m$ are the respective power-law coefficients. To solve the proposed viscoelastic model in its dimensional form, we will need to define six parameters ($T_0, \beta, m, n, \eta_0$ and $\tau_0$) and an initial condition on $\lambda$. We non-dimensionalize the proposed model by using the following variables: $t = \bar{t}/T_0$, $\sigma = (\beta T_0/\eta_0)\bar{\sigma}$, $\dot{\gamma} = \beta T_0\bar{\dot{\gamma}}$, $\eta =$ $\bar{\eta}/\eta_0$, and $\tau = \bar{\tau}/\tau_0$. Consequently, Eqs. (8) to (11) in their non-dimensional form can be rewritten as,

$$\frac{d\lambda}{dt} = 1 - \frac{\lambda}{\eta(\lambda)}\sigma \tag{12}$$

$$\frac{\tau_0}{T_0}\frac{d}{dt}\left(\frac{\tau(\lambda)}{\eta(\lambda)}\sigma\right) + \frac{\sigma}{\eta(\lambda)} = \dot{\gamma}_e + \dot{\gamma}_v = \dot{\gamma} \tag{13}$$

$$\eta(\lambda) = (1 + \lambda^n), and \tag{14}$$

$$\tau(\lambda) = (1 + \lambda^m) \tag{15}$$

In the structural kinetic modeling approach, the time-dependent evolution of rheological properties, such as viscosity, relaxation time, and elastic moduli, is governed by the structural parameter $\lambda$, which evolves under the competing influences of aging and rejuvenation. This evolution directly controls the material's response to applied deformation fields like shear or creep.

In creep experiment (constant stress or viscosity bifurcation experiment), the evolution of the structure parameter and the corresponding shear rate are determined by numerically solving the Eqs. (12) and (13), interestingly, their steady state forms (represented using subscript "ss") allow analytical treatment. In the Maxwell model, in the limit of steady state, the elastic strain in the spring remains constant, and hence the corresponding shear rate becomes zero ($\dot{\gamma}_{e,ss} = 0$). Consequently, only the shear rate associated with the viscous dashpot is present in a material $\dot{\gamma}_{ss} = \dot{\gamma}_{v,ss}$, which, when combined with Eq. 12, leads to the steady-state structure parameter ($\lambda_{ss}$) given by $\lambda_{ss} = 1/(\dot{\gamma}_{ss})$. Furthermore, from Eqs. (12) and (13) we get the relationship between the steady-state shear rate and the shear stress as,

$$\sigma_{ss} = (1 + (\dot{\gamma}_{ss})^{-n})\,\dot{\gamma}_{ss}. \tag{16}$$

This relation predicts a non-monotonic flow curve for hyper-aging systems where $n > 1$. In such cases, below a critical shear rate, the steady-state stress decreases as the shear rate increases. This behavior, as reported in the literature, is a consequence of the time-dependent increase in yield stress of hyper-aging systems.[23, 76]

Furthermore, to understand how static yield stress and yield strain evolve, consider a sample that has been pre-sheared at a shear rate $\dot{\gamma}_{PS}$ (in our case 500 s$^{-1}$). Immediately after the cessation of flow at $t = 0$, according to Eq. (12), the structural parameter begins to evolve as $\lambda = \lambda_{PS} + t$ (where $\lambda_{PS}$ is the steady state value of the structure parameter corresponding to the pre-shear $\dot{\gamma}_{PS}$). The system reaches a critical state at $\lambda = \lambda_c = 1/\dot{\gamma}_c$ after a time $t = t_c$. For applied shear stresses ($\sigma$) below the critical shear stress ($\sigma_c$), the shear rate induced in the system diminishes over time, and eventually, as shown in Fig. 9, the flow stops. Consequently, the minimum stress associated with the steady-state flow curve, i.e. $\sigma_c$ can be considered as the minimum yield stress associated with the system, which remains constant till $t \leq t_c$ (Eq. 17). During this period, however, the elastic modulus increases as $G' = \bar{G}'/G_0 = \lambda^{n-\mu} \sim t^{(n-\mu)}$, which results in a decrease in the yield strain over time given by Eq. (18).

$$\sigma_y = \sigma_c \qquad\qquad for\ \lambda \leq \lambda_c\ (or\ t \leq t_c) \qquad (17)$$

$$\gamma_y = \frac{\sigma_c}{G'(t; t \leq t_c)} = \frac{\sigma_c}{t^{(n-\mu)}} \qquad for\ \lambda \leq \lambda_c\ (or\ t \leq t_c) \qquad (18)$$

Beyond the critical aging time ($t > t_c$), the structure parameter exceeds the critical threshold ($\lambda > \lambda_c$). In this regime, whether the material flows or arrests depends on the sign of $d\lambda/dt$. If the applied stress is insufficient to rejuvenate the system, $d\lambda/dt > 0$, aging dominates, viscosity continues to increase, and the shear rate progressively decreases to zero. On the other hand, if applied stress is large enough to overcome aging, then $d\lambda/dt < 0$, leading to a stable steady flow. In this $t > t_c$ regime, the yield stress and strain evolve as:

$$\sigma_y = \left[\frac{(1 + (\lambda_{PS} + t)^n)}{\lambda_{PS} + t}\right] \qquad for\ \lambda > \lambda_c\ (or\ t > t_c) \qquad (19)$$

$$\gamma_y = \frac{1}{t^{(n-\mu)}}\left[\frac{(1 + (\lambda_{PS} + t)^n)}{\lambda_{PS} + t}\right] \qquad for\ \lambda > \lambda_c\ (or\ t > t_c) \qquad (20)$$

Notably, from Eqs. (19) and (20), for hyper-aging materials ($\mu > 1$), both the yield stress and strain increase with time for $t > t_c$.

In order to numerically solve Eqs. (12) and (13) for the constant stress (viscosity bifurcation) flow field, fit the steady-state flow curve, and time-dependent yield stress/strain, accurate values of model parameters $m, n, \eta_0, \tau_0, \beta$, and $T_0$ must be determined. Interestingly, all these parameters can be obtained from the experimental measurements as follows.

- Aging exponents ($m, n$): In the absence of flow (quiescent aging), the rejuvenation term in Eq. (12) vanishes, and $\lambda = \lambda_{PS} + t$. At long times, for $t \gg \lambda_{PS}$, and $\lambda \gg 1$, Eqs. (14) and (15) lead to, $\tau(t) \sim t^m$ and $G'(t) = \eta(\lambda)/\tau(\lambda) \sim t^{n-m}$. However, the time aging time superposition (Fig. 7 (b) and 8) has shown that $\tau(t) \sim t^\mu$, while Fig. 4 suggests $G' = G_0 t^q$ in a limit of long time. Thus, for $t \gg \lambda_{PS}$, and $t \gg 1$, we get $m = \mu$ and $n \approx \mu + q$, as well as $G_0$.
- Viscosity ($\eta_0$) and relaxation time ($\tau_0$) of the system in a structure-less state: The parameter $\eta_0$ is determined by fitting the steady-state viscosity data to the empirical equation $\eta = \eta_0 - a\dot{\gamma}_{ss}^{-b}$, where $\eta \to \eta_0$ as $\dot{\gamma}_{ss} \to \infty$. Furthermore, $\tau_0 = \eta_0/G_0$.
- Parameters $\beta$ and $T_0$: Differentiating the non-monotonic steady-state flow curve Eq. (16) and setting $\left(\frac{d\sigma_{ss}}{d\dot{\gamma}_{ss}}\right)_{\dot{\gamma}_{ss}=\dot{\gamma}_c} = 0$, we get the relationship between $\beta T_0$ and experimental parameters $n$ and $\dot{\gamma}_c$ as,

$$\beta T_0 = \frac{(n-1)^{1/n}}{\dot{\gamma}_c} \tag{21}$$

From this, $\beta T_0$ is calculated. Using $\lambda_{PS} = 1/\beta T_0 \dot{\gamma}_{PS}$, and $\lambda_c = 1/\beta T_0 \dot{\gamma}_c$, and knowing that the yield stress remains constant up to time $t_c = 100$ s (Fig. 12 (a)), $T_0$ is calculated as: $T_0 = t_c / (\lambda_c - \lambda_{PS})$.

This systematic estimation of model parameters enables us to explore the macroscopic rheological behavior of silica gel through the lens of structural kinetic formalism. By integrating parameters directly obtained from experimental measurement, the model serves as a useful framework for interpreting the evolution of flow and aging behavior of the system under different deformation conditions.

We begin by examining the time-dependent evolution of the model under constant stress, wheerin Eqs. (12) and (13) are solved simultaneously for stresses ($\sigma$) above and below the critical value $\sigma_c = 1.901$. As shown in Fig. 13 (a), for stresses below $\sigma_c$, the structure parameter $\lambda$ increases progressively with time, indicating continuous aging. At lower stresses, this increase starts early and progresses gradually, whereas, for stresses just below $\sigma_c$, the structure parameter $\lambda$ initially increases weakly at intermediate times before rising steeply. At larger times, $\lambda \sim t$, indicating that for $\sigma < \sigma_c$, the system's structure grows continuously, leading to flow cessation. For $\sigma \geq \sigma_c$, however, the system attains a steady-state structural parameter and flows homogeneously, reflecting a stable balance between aging and rejuvenation. The corresponding evolution of the shear rate is shown in Fig. 13(b). For $\sigma < \sigma_c$, the shear rate gradually diminishes and eventually approaches zero. In contrast, for $\sigma \geq \sigma_c$, the system achieves a finite, steady shear rate. These trends are consistent with experimental observations and highlight how the model captures the essential features of the flow-to-arrest transition above and below the critical shear stress ($\sigma_c$) seen in the viscosity bifurcation experiment, and correlates it to the structural evolution in the soft glassy materials.

While the single-parameter model successfully captures the essential interplay between aging and rejuvenation and provides a coherent explanation for the observed transition between flow (finite constant viscosity) and arrest (infinite viscosity), some finer features in the experimental shear rate evolution – such as the non-monotonic evolution of the shear rate observed in Fig. 9 (wherein for $\sigma < \sigma_c$, the shear rate initially increases and then starts to decrease) are not reproduced. This discrepancy arises because the single structural variable $\lambda$, which evolves monotonically, cannot capture the non-monotonicity seen in the experiment. Although not the focus of the present study, similar to the multi-mode approach used in the viscoelasticity modelling, the multi-mode structural kinetic model with different $\lambda_i$ (with $i = 1, 2 \ldots$ as thixotropic modes) may better capture such complexities by accounting for the structural evolution happening at different length and time scales.[83]

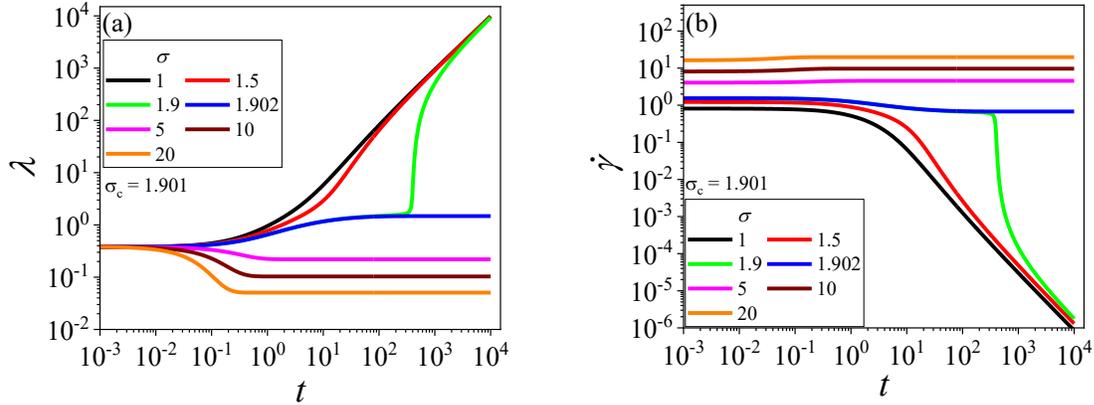

**Figure 13:** (a) Time evolution of structure parameter ($\lambda$) for various applied stresses. (b) Corresponding time evolution of shear rate ($\dot{\gamma}$) with time for the same stresses. The model parameters used are $n = 1.52$, $\mu = 1.20$, $\tau_0 = 1.18 \times 10^{-4}$ s, and $T_0 = 77.94$ s, yielding a critical stress ($\sigma_c$) of 1.901. The values of the different applied stresses (below and above the yield stress) are given in the legend.

Interestingly, as shown in Fig. 14, the model effectively captures the non-monotonic behavior of the steady-state flow curve observed in hyper-aging systems like silica dispersion. In these systems, below the critical shear rate ($\dot{\gamma}_c$), the flow becomes unstable, with stress decreasing as shear rate increases. Although this unstable regime is experimentally inaccessible with a stress-controlled rheometer, the structural kinetic model successfully captures the non-monotonic flow curve behavior. In contrast, for systems exhibiting simple aging ($n = 1$), the model yields a monotonic flow curve characterized by a yield stress plateau at low shear rates, consistent with experimental observations in micro-gel pastes and hard-sphere glasses, wherein the yield stress of the system remains constant over time.[23] Conversely, in hyper-aging silica dispersions, where the yield stress evolves with time, non-monotonic flow curves have been observed.[23]

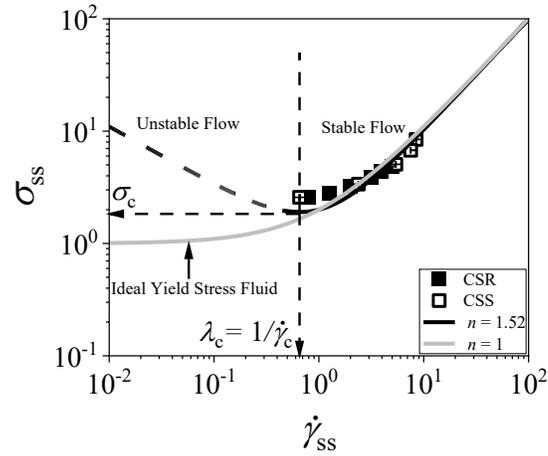

**Figure 14:** Non-dimensional shear stress versus shear rate steady-state flow curve for 1-day-old silica dispersion (CSR: Constant shear rate (closed symbol), CSS: Constant shear stress (open symbol)), fitted flow curve of the structural kinetic model for hyper-aging ($n = 1.52$), and prediction for simple aging ($n = 1$) systems. In model fit, the stable part of the flow curve for the hyper-aging system is shown as a black solid line, whereas the unstable part is shown by a black dashed line. The model prediction for a simple aging system is shown by a gray line. For non-dimensionalizing the experimental shear stress and shear rate, the values of model parameters used are: $\beta = 1.01 \times 10^{-4}$, $T_0 = 77.94$ s, and $\eta_0 = 9.38 \times 10^{-3}$ Pa.s.

To further understand the time-dependent yielding transition in the hyper-aging silica dispersion, we compare the model's output with the experimental evolution of yield stress and yield strain, as shown in Fig. 15 (a) and 15 (b). As seen in Fig. 15 (a), the model captures the trend very well, wherein the yield stress remains constant up to the critical aging time $t_c$, and then increases with time, consistent with experimental observation. Similarly, the model qualitatively reproduces the evolution of yield strain in Fig. 15(b): initially decreasing with time due to structural stiffening (increase in modulus), followed by a gradual increase at later times.

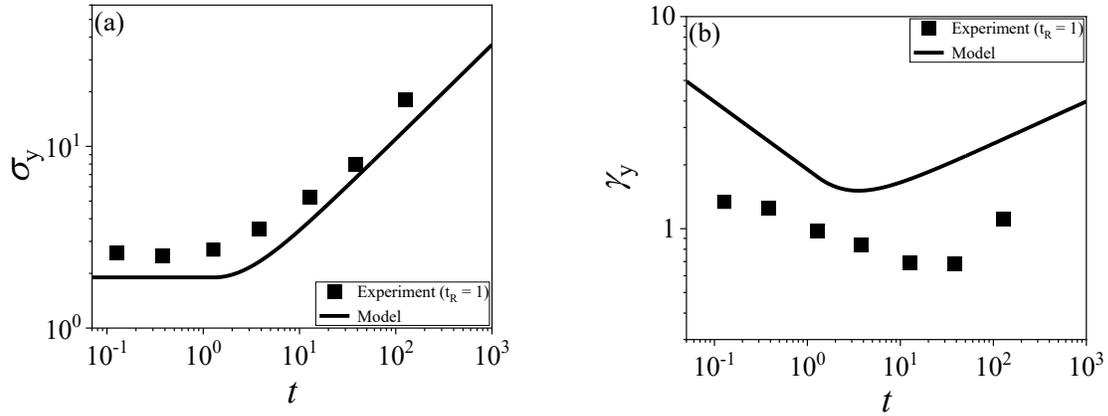

**Figure 15:** Experimental results (solid symbol) and model fit (solid line) for the evolution of (a) yield stress ($\sigma_y$), and (b) yield strain ($\gamma_y$) with time for hyper-aging colloidal silica dispersion ($n = 1.52, \mu = 1.203$). For non-dimensionalizing the experimental yield stress and yield strain, the values of model parameters used are: $\beta = 1.01 \times 10^{-4}$, $T_0 = 77.94$ s, $\eta_0 = 9.38 \times 10^{-3}$ Pa.s, and $G_0 = 79.51$ Pa.

Overall, the structural kinetic model gives us keen insight into the rheological behavior of the hyper-aging colloidal silica gel. It effectively captures the key rheological responses such as the non-monotonic flow curve, viscosity bifurcation, and time-dependent yield stress and yield strain. While certain quantitative discrepancies between model predictions and experimental results exist, these can be reasonably attributed to two key simplifications. First, the use of a single structure parameter $\lambda$ to represent the system's instantaneous structural state, though it effectively captures the overall behavior, it also averages out the effect of the structural evolution happening at different lengths and time scales. Second, the time-independent values of power law coefficients $\mu, q$, as well as the magnitude of elastic modulus in the structureless state ($G_0$) considered for solving the model, might not fully represent the dynamic changes occurring over the entire experimental duration. As shown in Fig. 4, the $G'$ shows a faster growth at smaller $t_w$ (smaller $G_0$, larger $q$) compared to larger $t_w$ (larger $G_0$, smaller $q$); similarly, the relaxation time also may grow faster at smaller $t_w$ than at larger $t_w$. Despite its limitation in capturing quantitative behavior, the model's qualitative agreement with the experimentally observed trend highlights its effectiveness in describing the thixotropic behavior of colloidal silica dispersion and capturing the underlying physics by relating the evolution of the structure parameter to the material response under different flow conditions.

The natural next step in our modeling effort would be to extend the structural kinetic formalism to rheological data obtained at different rest times. However, as highlighted in the preceding discussion, the structural kinetic model employed in this work consists of only a single viscoelastic relaxation mode along with a single $\lambda$ component, leading to a single thixotropic timescale. Moreover, in our current approach, we make considerable simplifying assumptions regarding the evolution of both the elastic modulus and the relaxation time, particularly during the initial phase subsequent to cessation of shear melting. Shear melting after the rest time introduces nontrivial effects on the rheological response of colloidal Ludox silica dispersions. Specifically, we fathom that not just the relaxation time spectrum of the dispersion gets altered, but the spectrum of the thixotropic timescales that govern structural evolution also gets significantly impacted. As a result, the aging/rejuvenation dynamics of the system become increasingly complex with rest time, making the assumption of single-mode behavior progressively less tenable. Given this complexity, applying the proposed single-mode Maxwell-based structural kinetic model uniformly across different rest times is unlikely to yield physically insightful results. Therefore, instead of pursuing a potentially misleading extension with the current framework, we choose to carry out a more rigorous modeling of the rest-time-dependent rheological behavior as future work. Such an effort would necessarily require a more sophisticated model that incorporates both a multimodal viscoelastic spectrum and a multimodal structural (thixotropic) relaxation spectrum as discussed elsewhere.[83]

Finally, we discuss the origin of physical aging in Ludox silica dispersions, which is observed to be reversible over short time scales (on the order of hours) but becomes irreversible over longer durations (days). As a result, shear melting performed after a long rest period is no longer capable of fully restoring the system to its freshly prepared state. As mentioned earlier, these silica particles carry a negative charge when dispersed in aqueous media.[38] The presence of NaCl leads to ion dissociation, which screens the electrostatic repulsion between the particles, disrupts hydrogen bonding between the silica surfaces and water molecules, and enables closer approach of the particles.[6, 39, 40] Furthermore, in the presence of salt, surface heterogeneity also leads to the formation of patchy charged domains on silica surfaces, giving rise to weak attractive interactions even between like-charged particles.[44] These aggregates then associate via van der Waals interactions to form a percolated gel network. This structure is out of thermodynamic equilibrium and therefore a process of physical aging gets initiated in the same, during which the silica particles undergo microscopic structural rearrangements while remaining part of the network. These rearrangements drive the system to progressively lower energy states, resulting in a gradual decrease in free energy. This progressive lowering of free energy involves both strengthening of interparticle bonds and densification of the network, leading to an increase in the elastic modulus over time. When a deformation field is applied, these newly formed bonds break, leading to shear melting or rejuvenation of the Ludox gel system. This process increases the free energy of the system and effectively reverses the structure formed during physical aging. The Ludox gel system thus exhibits several characteristic features of soft glassy materials, including a non-monotonic flow curve, time-dependent yield stress and yield strain, and viscosity bifurcation. Furthermore, since the rate of microscopic structural rearrangement increases with temperature, the evolution of dynamic moduli shifts to shorter times at elevated temperatures.

As the system ages over several days, the interparticle bonds between silica particles strengthen to such an extent that the applied deformation field can no longer break them completely. This increasing resistance to shear rejuvenation arises from the growth and stabilization of covalently bonded non-Brownian aggregates. With time, due to close proximity, more silica particles participate in siloxane bond $(\mathrm{Si-O-Si})$ formation, resulting in the formation of larger non-Brownian aggregates.[41-43] When the shear melting is performed, the weaker van der Waals bonds between the non-Brownian aggregates get readily disrupted. Upon cessation of shear, these weak bonds quickly reform the gel network through their faster kinetics. In contrast, the formation of additional siloxane bonds occurs over much longer timescales (days). However, application of shear cannot break these stronger bonds. This inability of shear rejuvenation to break stronger bonds leads to irreversible aging. Consequently, the steady-state shear viscosity during shear melting progressively rises with increasing rest time. In subsequent aging experiments, because the process begins from a more matured (lower energy) state with a higher fraction of unbroken aggregates, the overall evolution shifts to lower times for samples with higher rest time. This inherent unbroken structure is reflected in flow curves shifting to higher stress levels and increased values of yield stress. In general, we observe that the relaxation time during aging increases more strongly than linearly with time ($\tau \sim t^\mu$, $\mu > 1$), indicating hyper-aging dynamics. Such behavior may stem from enthalpic aging driven by interparticle attractive interactions and associated structural arrest, wherein progressive bond formation leads to the deepening of energy wells, as discussed elsewhere.[56, 76, 84] However, since shear melting performed at later stages does not completely disrupt the aggregates, the contribution of the enthalpic component to aging gradually diminishes. We believe this is responsible for the observed decrease in the power-law exponent $\mu$ with increasing rest time, as the system approaches a regime of restricted caged dynamics, where further structural rearrangements become progressively limited. Eventually, at very long rest times (on the order of several months), $\mu$ approaches unity, which corresponds to the limiting case of simple aging.

**Conclusion**

This study reveals the complex rheological behavior of colloidal Ludox TM-40 silica dispersions, which is comprised of the space-spanning percolated network of silica nanoparticles, giving it a soft-solid like consistency. Through a series of time- and deformation-field-dependent experiments, we investigate its physical aging and rejuvenation behavior characterized by the evolution of its viscoelastic properties as a function of time (in days) elapsed since the preparation of the silica dispersion, termed as rest time. We have shown that these gels exhibit hallmarks of soft glassy behavior, including an increase in elastic modulus and relaxation time with age, viscosity bifurcation with a well-defined critical stress, and time-dependent yield stress and strain. These phenomena point to enthalpy-driven structural evolution and activated microscopic rearrangements within the gel. Interestingly, while the short-term aging in the colloidal silica gel system can be reversed by deformation (shear rejuvenation), the longer rest times lead to irreversible aging. This irreversibility stems from the formation of stronger, shear-resistant interparticle siloxane bonds that cannot be fully broken down, resulting in increased steady-state viscosity, higher minimum yield stress, and an increase in the critical shear rate required for the homogeneous flow over time.

Furthermore, we also successfully model the key rheological features, such as time-dependent yield stress and strain, viscosity bifurcation, and non-monotonic flow curves, using a simple single-parameter time-dependent Maxwell model in the structural kinetic model formalism. We believe that a more sophisticated framework incorporating multiscale structural evolution and multimodel viscoelastic formulation could fully capture the rest-time-dependent changes. Our findings on irreversible aging dynamics and its consequences on the macroscopic rheological response of the colloidal silica gel are very important for optimizing formulations and predicting the performance of silica-based materials in various industries, including construction, pharmaceuticals, agrochemicals, home and personal care, energy storage, etc., where processing history and storage conditions significantly impact product quality. Therefore, the insights from this work can aid in optimizing the formulation strategies and tailoring the time-dependent rheology of the silica-based systems to meet the specific functional demands.

**Supporting Information Available**

Particle size distribution histogram, Time – aging-time superposition, evolution of shear rate with time at different applied stresses (0.5 – 10 Pa), amplitude sweep plots for determining the static yield stress for 6, 12, 18, and 24 days old silica dispersion, and a list of symbols.

**Acknowledgment**

YMJ would like to acknowledge financial support from the Science and Engineering Research Board, Government of India (Grant Nos. CRG/2022/004868 and JCB/2022/000040). We also thank the "*Centre for Nanosciences at IIT Kanpur*" for the FESEM imaging.

**Author Declarations:**

The authors have no conflicts to disclose.

**References**

(1) Larson, R. G. *The structure and rheology of complex fluids*; Oxford university press New York, 1999.

(2) Stark, W. J.; Stoessel, P. R.; Wohlleben, W.; Hafner, A. Industrial applications of nanoparticles. *Chemical Society Reviews* **2015**, *44* (16), 5793-5805.

(3) Slowing, I. I.; Trewyn, B. G.; Giri, S.; Lin, V. Y. Mesoporous silica nanoparticles for drug delivery and biosensing applications. *Advanced functional materials* **2007**, *17* (8), 1225-1236.

(4) Jiang, X.; Yang, F.; Jia, W.; Jiang, Y.; Wu, X.; Song, S.; Shen, H.; Shen, J. Nanomaterials and nanotechnology in agricultural pesticide delivery: A review. *Langmuir* **2024**, *40* (36), 18806-18820.

(5) Teoman, B.; Potanin, A.; Armenante, P. M. Optimization of optical transparency of personal care products using the refractive index matching method. *Colloids and Surfaces A: Physicochemical and Engineering Aspects* **2021**, *610*, 125595.

(6) Sögaard, C.; Funehag, J.; Abbas, Z. Silica sol as grouting material: a physio-chemical analysis. *Nano convergence* **2018**, *5*, 1-15.

(7) Funehag, J.; Fransson, Å. Sealing narrow fractures with a Newtonian fluid: model prediction for grouting verified by field study. *Tunnelling and underground space technology* **2006**, *21* (5), 492-498.

(8) Horowitz, A. I.; Panzer, M. J. High-performance, mechanically compliant silica-based ionogels for electrical energy storage applications. *Journal of Materials Chemistry* **2012**, *22* (32), 16534-16539.

(9) Joshi, Y. M. Dynamics of colloidal glasses and gels. *Annual review of chemical and biomolecular engineering* **2014**, *5* (1), 181-202.

(10) Müller, F. J.; Ramakrishna, S. N.; Isa, L.; Vermant, J. Tuning Colloidal Gel Properties: The Influence of Central and Noncentral Forces. *Langmuir* **2025**.

(11) Marium, M.; Hoque, M.; Miran, M. S.; Thomas, M. L.; Kawamura, I.; Ueno, K.; Dokko, K.; Watanabe, M. Rheological and ionic transport properties of nanocomposite electrolytes based on protic ionic liquids and silica nanoparticles. *Langmuir* **2019**, *36* (1), 148-158.

(12) Like, B. D.; Panzer, M. J. Sol–Gel Synthesis of Phosphorylcholine Zwitterion-Decorated Silica Gels. *Langmuir* **2025**.

(13) Hom, W. L.; Bhatia, S. R. Significant enhancement of elasticity in alginate-clay nanocomposite hydrogels with PEO-PPO-PEO copolymers. *Polymer* **2017**, *109*, 170-175.

(14) Liu, X.; Bhatia, S. R. Laponite® and Laponite®-PEO hydrogels with enhanced elasticity in phosphate-buffered saline. *Polymers for Advanced Technologies* **2015**, *26* (7), 874-879.

(15) Kishore, S.; Srivastava, S.; Bhatia, S. R. Microstructure of colloid-polymer mixtures containing charged colloidal disks and weakly-adsorbing polymers. *Polymer* **2016**, *105*, 461-471.

(16) Cipelletti, L.; Ramos, L. Slow dynamics in glassy soft matter. *Journal of Physics: Condensed Matter* **2005**, *17* (6), R253.

(17) Hunter, G. L.; Weeks, E. R. The physics of the colloidal glass transition. *Reports on progress in physics* **2012**, *75* (6), 066501.

(18) Petekidis, G.; Wagner, N. J.; Mewis, J. Rheology of colloidal glasses and gels. *Theory and Applications of Colloidal Suspension Rheology* **2021**, 173-226.

(19) Mewis, J.; Wagner, N. J. Thixotropy. *Advances in colloid and interface science* **2009**, *147*, 214-227.

(20) Agarwal, M.; Sharma, S.; Shankar, V.; Joshi, Y. M. Distinguishing thixotropy from viscoelasticity. *Journal of Rheology* **2021**, *65* (4), 663-680.

(21) Coussot, P.; Nguyen, Q. D.; Huynh, H.; Bonn, D. Viscosity bifurcation in thixotropic, yielding fluids. *Journal of rheology* **2002**, *46* (3), 573-589.

(22) Fall, A.; Paredes, J.; Bonn, D. Yielding and shear banding in soft glassy materials. *Physical review letters* **2010**, *105* (22), 225502.

(23) Bhattacharyya, T.; Jacob, A. R.; Petekidis, G.; Joshi, Y. M. On the nature of flow curve and categorization of thixotropic yield stress materials. *Journal of Rheology* **2023**, *67* (2), 461-477.

(24) Divoux, T.; Agoritsas, E.; Aime, S.; Barentin, C.; Barrat, J.-L.; Benzi, R.; Berthier, L.; Bi, D.; Biroli, G.; Bonn, D. Ductile-to-brittle transition and yielding in soft amorphous materials: perspectives and open questions. *Soft Matter* **2024**, *20* (35), 6868-6888.

(25) Kamani, K. M.; Rogers, S. A. Brittle and ductile yielding in soft materials. *Proceedings of the National Academy of Sciences* **2024**, *121* (22), e2401409121.

(26) Kurokawa, A.; Vidal, V.; Kurita, K.; Divoux, T.; Manneville, S. Avalanche-like fluidization of a non-Brownian particle gel. *Soft Matter* **2015**, *11* (46), 9026-9037.

(27) Vasu, K.; Krishnaswamy, R.; Sampath, S.; Sood, A. Yield stress, thixotropy and shear banding in a dilute aqueous suspension of few layer graphene oxide platelets. *Soft Matter* **2013**, *9* (25), 5874-5882.

(28) Agarwal, M.; Kaushal, M.; Joshi, Y. M. Signatures of overaging in an aqueous dispersion of carbopol. *Langmuir* **2020**, *36* (48), 14849-14863.

(29) Viasnoff, V.; Lequeux, F. Rejuvenation and overaging in a colloidal glass under shear. *Physical Review Letters* **2002**, *89* (6), 065701.

(30) Sudreau, I.; Auxois, M.; Servel, M.; Lécolier, É.; Manneville, S.; Divoux, T. Residual stresses and shear-induced overaging in boehmite gels. *Physical Review Materials* **2022**, *6* (4), L042601.

(31) Joshi, Y. M. Thixotropy, nonmonotonic stress relaxation, and the second law of thermodynamics. *Journal of Rheology* **2022**, *66* (1), 111-123.

(32) Joshi, Y. M.; Shahin, A.; Cates, M. E. Delayed solidification of soft glasses: new experiments, and a theoretical challenge. *Faraday discussions* **2012**, *158* (1), 313-324.

(33) Ilyin, S.; Arinina, M.; Malkin, A. Y.; Kulichikhin, V. Sol–gel transition and rheological properties of silica nanoparticle dispersions. *Colloid Journal* **2016**, *78*, 608-615.

(34) Kumar, V.; McKinley, G. H.; Joshi, Y. M. Shear Stress Build-up Under Constant Strain Conditions in Soft Glassy Materials. *arXiv preprint arXiv:2504.00806* **2025**.

(35) Iler, R. The chemistry of silica, A Wiley-Interscience publication. *Willey and Sons, New York* **1979**, 665-676.

(36) Sögaard, C.; Hagström, M.; Abbas, Z. Temperature and Particle-size Effects on the Formation of Silica Gels from Silica Sols. *Silicon* **2023**, *15* (8), 3441-3451.

(37) Simonsson, I.; Sögaard, C.; Rambaran, M.; Abbas, Z. The specific co-ion effect on gelling and surface charging of silica nanoparticles: Speculation or reality? *Colloids and Surfaces A: Physicochemical and Engineering Aspects* **2018**, *559*, 334-341.

(38) Kosmulski, M. The pH dependent surface charging and points of zero charge. VIII. Update. *Advances in Colloid and Interface Science* **2020**, *275*, 102064.

(39) Allen, L. H.; Matijevic, E. Stability of colloidal silica: II. Ion exchange. *Journal of colloid and interface science* **1970**, *33* (3), 420-429.

(40) Allen, L. H.; Matijević, E. Stability of colloidal silica: I. Effect of simple electrolytes. *Journal of colloid and interface science* **1969**, *31* (3), 287-296.

(41) Depasse, J.; Watillon, A. The stability of amorphous colloidal silica. *Journal of colloid and interface science* **1970**, *33* (3), 430-438.

(42) Depasse, J. Coagulation of colloidal silica by alkaline cations: Surface dehydration or interparticle bridging? Elsevier: 1997; Vol. 194, pp 260-262.

(43) Drabarek, E.; Bartlett, J. R.; Hanley, H.; Woolfrey, J.; Muzny, C. Effect of processing variables on the structural evolution of silica gels. *International journal of thermophysics* **2002**, *23*, 145-160.

(44) Chen, W.; Tan, S.; Zhou, Y.; Ng, T.-K.; Ford, W. T.; Tong, P. Attraction between weakly charged silica spheres at a water-air interface induced by surface-charge heterogeneity. *Physical Review E—Statistical, Nonlinear, and Soft Matter Physics* **2009**, *79* (4), 041403.

(45) Cao, X.; Cummins, H.; Morris, J. Structural and rheological evolution of silica nanoparticle gels. *Soft Matter* **2010**, *6* (21), 5425-5433.

(46) Metin, C. O.; Rankin, K. M.; Nguyen, Q. P. Phase behavior and rheological characterization of silica nanoparticle gel. *Applied Nanoscience* **2014**, *4*, 93-101.

(47) Hatami, S.; Hughes, T. J.; Sun, H.; Roshan, H.; Walsh, S. D. On the application of silica gel for mitigating CO2 leakage in CCS projects: Rheological properties and chemical stability. *Journal of Petroleum Science and Engineering* **2021**, *207*, 109155.

(48) Suman, K.; Mittal, M.; Joshi, Y. M. Effect of sodium pyrophosphate and understanding microstructure of aqueous LAPONITE® dispersion using dissolution study. *Journal of Physics: Condensed Matter* **2020**, *32* (22), 224002.

(49) Manley, S.; Davidovitch, B.; Davies, N. R.; Cipelletti, L.; Bailey, A.; Christianson, R. J.; Gasser, U.; Prasad, V.; Segre, P.; Doherty, M. Time-dependent strength of colloidal gels. *Physical review letters* **2005**, *95* (4), 048302.

(50) Okazaki, K.; Kawaguchi, M. Influence of monovalent electrolytes on rheological properties of gelled colloidal silica suspensions. *Journal of dispersion science and technology* **2008**, *29* (1), 77-82.

(51) Møller, P.; Rodts, S.; Michels, M.; Bonn, D. Shear banding and yield stress in soft glassy materials. *Physical Review E—Statistical, Nonlinear, and Soft Matter Physics* **2008**, *77* (4), 041507.

(52) Ghaffari, Z.; Rezvani, H.; Khalilnezhad, A.; Cortes, F. B.; Riazi, M. Experimental characterization of colloidal silica gel for water conformance control in oil reservoirs. *Scientific Reports* **2022**, *12* (1), 9628.

(53) Ianni, F.; Di Leonardo, R.; Gentilini, S.; Ruocco, G. Aging after shear rejuvenation in a soft glassy colloidal suspension: Evidence for two different regimes. *Physical Review E— Statistical, Nonlinear, and Soft Matter Physics* **2007**, *75* (1), 011408.

(54) Bonn, D.; Tanaka, H.; Coussot, P.; Meunier, J. Ageing, shear rejuvenation and avalanches in soft glassy materials. *Journal of Physics: Condensed Matter* **2004**, *16* (42), S4987.

(55) Mewis, J.; Wagner, N. J. *Colloidal suspension rheology*; Cambridge university press Cambridge, 2012.

(56) Joshi, Y. M. Linear Viscoelasticity of Physically Aging Soft Glassy (Thixotropic) Materials. *Current Opinion in Colloid & Interface Science* **2025**, 101896.

(57) Israelachvili, J. N. *Intermolecular and surface forces*; Academic press, 2011.

(58) Kaushal, M.; Joshi, Y. M. Linear viscoelasticity of soft glassy materials. *Soft Matter* **2014**, *10* (12), 1891-1894, 10.1039/C3SM52978A. DOI: 10.1039/C3SM52978A.

(59) Kumar, A.; Kumar, V.; Joshi, Y. M.; Singh, M. K. Tribological and Rheological Study of Thixotropic Gels of 2D Nanoparticles. *Langmuir* **2024**, *40* (14), 7310-7327.

(60) Agarwal, M.; Joshi, Y. M. Signatures of physical aging and thixotropy in aqueous dispersion of Carbopol. *Physics of Fluids* **2019**, *31* (6).

(61) Ferry, J. *Viscoelastic Properties of Polymers*; Wiley, 1980.

(62) Fielding, S. M.; Sollich, P.; Cates, M. E. Aging and rheology in soft materials. *Journal of Rheology* **2000**, *44* (2), 323-369.

(63) Malkin, A. Y.; Isayev, A. I. *Rheology: concepts, methods, and applications*; Elsevier, 2022.

(64) Struik, L. C. E. *Physical aging in amorphous polymers and other materials*; Citeseer, 1978.

(65) Shahin, A.; Joshi, Y. M. Prediction of long and short time rheological behavior in soft glassy materials. *Physical Review Letters* **2011**, *106* (3), 038302.

(66) Rathinaraj, J. D. J.; Lennon, K. R.; Gonzalez, M.; Santra, A.; Swan, J. W.; McKinley, G. H. Elastoviscoplasticity, hyperaging, and time–age-time–temperature superposition in aqueous dispersions of bentonite clay. *Soft Matter* **2023**, *19* (38), 7293-7312.

(67) Cloitre, M.; Borrega, R.; Leibler, L. Rheological aging and rejuvenation in microgel pastes. *Physical Review Letters* **2000**, *85* (22), 4819.

(68) Derec, C.; Ducouret, G.; Ajdari, A.; Lequeux, F. Aging and nonlinear rheology in suspensions of polyethylene oxide–protected silica particles. *Physical Review E* **2003**, *67* (6), 061403.

(69) Joshi, Y. M. Long time response of aging glassy polymers. *Rheologica Acta* **2014**, *53*, 477-488.

(70) Shahin, A.; Joshi, Y. M. Hyper-aging dynamics of nanoclay suspension. *Langmuir* **2012**, *28* (13), 5826-5833.

(71) Shahin, A.; Joshi, Y. M. Irreversible aging dynamics and generic phase behavior of aqueous suspensions of laponite. *Langmuir* **2010**, *26* (6), 4219-4225.

(72) Malkin, A.; Kulichikhin, V.; Ilyin, S. A modern look on yield stress fluids. *Rheologica Acta* **2017**, *56*, 177-188.

(73) Shoaib, M.; Bobicki, E. R. Rheological implications of pH induced particle–particle association in aqueous suspension of an anisotropic charged clay. *Soft Matter* **2021**, *17* (34), 7822-7834.

(74) Shoaib, M.; Molaei, N.; Bobicki, E. R. Physical aging in aqueous nematic gels of a swelling nanoclay: sol (phase) to gel (state) transition. *Physical Chemistry Chemical Physics* **2022**, *24* (8), 4703-4714.

(75) Shoaib, M.; Khan, S.; Wani, O. B.; Abdala, A.; Seiphoori, A.; Bobicki, E. R. Modulation of soft glassy dynamics in aqueous suspensions of an anisotropic charged swelling clay through pH adjustment. *Journal of Colloid and Interface Science* **2022**, *606*, 860-872.

(76) Joshi, Y. M. A model for aging under deformation field, residual stresses and strains in soft glassy materials. *Soft Matter* **2015**, *11* (16), 3198-3214.

(77) Barnes, H. A. The yield stress—a review or 'παντα ρει'—everything flows? *Journal of Non-Newtonian Fluid Mechanics* **1999**, *81* (1), 133-178. DOI: https://doi.org/10.1016/S0377-0257(98)00094-9.

(78) Shaukat, A.; Sharma, A.; Joshi, Y. M. Squeeze flow behavior of (soft glassy) thixotropic material. *Journal of Non-Newtonian Fluid Mechanics* **2012**, *167-168*, 9-17. DOI: https://doi.org/10.1016/j.jnnfm.2011.09.006.

(79) Negi, A. S.; Osuji, C. O. Time-resolved viscoelastic properties during structural arrest and aging of a colloidal glass. *Physical Review E—Statistical, Nonlinear, and Soft Matter Physics* **2010**, *82* (3), 031404.

(80) Moller, P.; Fall, A.; Chikkadi, V.; Derks, D.; Bonn, D. An attempt to categorize yield stress fluid behaviour. *Philosophical Transactions of the Royal Society A: Mathematical, Physical and Engineering Sciences* **2009**, *367* (1909), 5139-5155.

(81) Larson, R. Constitutive equations for thixotropic fluids. *Journal of Rheology* **2015**, *59* (3), 595-611.

(82) Joshi, Y. M. PERSPECTIVE: Analysis of thixotropic timescale. *Journal of Rheology* **2024**, *68* (4), 641-653.

(83) Wei, Y.; Solomon, M. J.; Larson, R. G. Modeling the nonmonotonic time-dependence of viscosity bifurcation in thixotropic yield-stress fluids. *Journal of Rheology* **2019**, *63* (4), 673-675.

(84) Joshi, Y. M.; Petekidis, G. Yield stress fluids and ageing. *Rheologica Acta* **2018**, *57*, 521-549.

**Rheological Behavior of Colloidal Silica Dispersion: Irreversible Aging and Thixotropy**

Vivek Kumar, Yogesh M. Joshi*

Department of Chemical Engineering, Indian Institute of Technology Kanpur, Kanpur, 208016, India

*Corresponding author

Email: joshi@iitk.ac.in

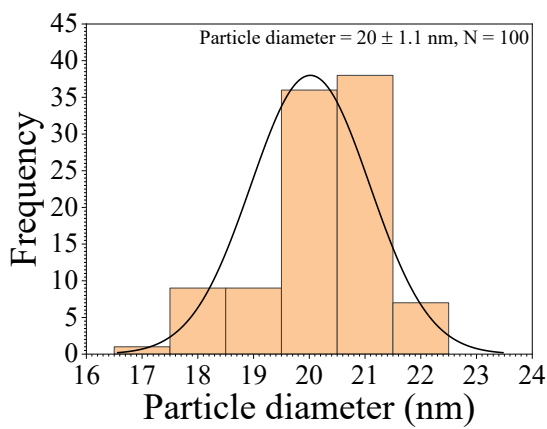

**Figure S1:** Histogram showing the size distribution of Ludox TM-40 colloidal silica particles as measured from FESEM image using ImageJ (N = 100). The data follow a near-Gaussian distribution with a mean particle diameter of 20 $\pm$ 1.1 nm. Bin width = 1 nm.

## Time aging time superposition

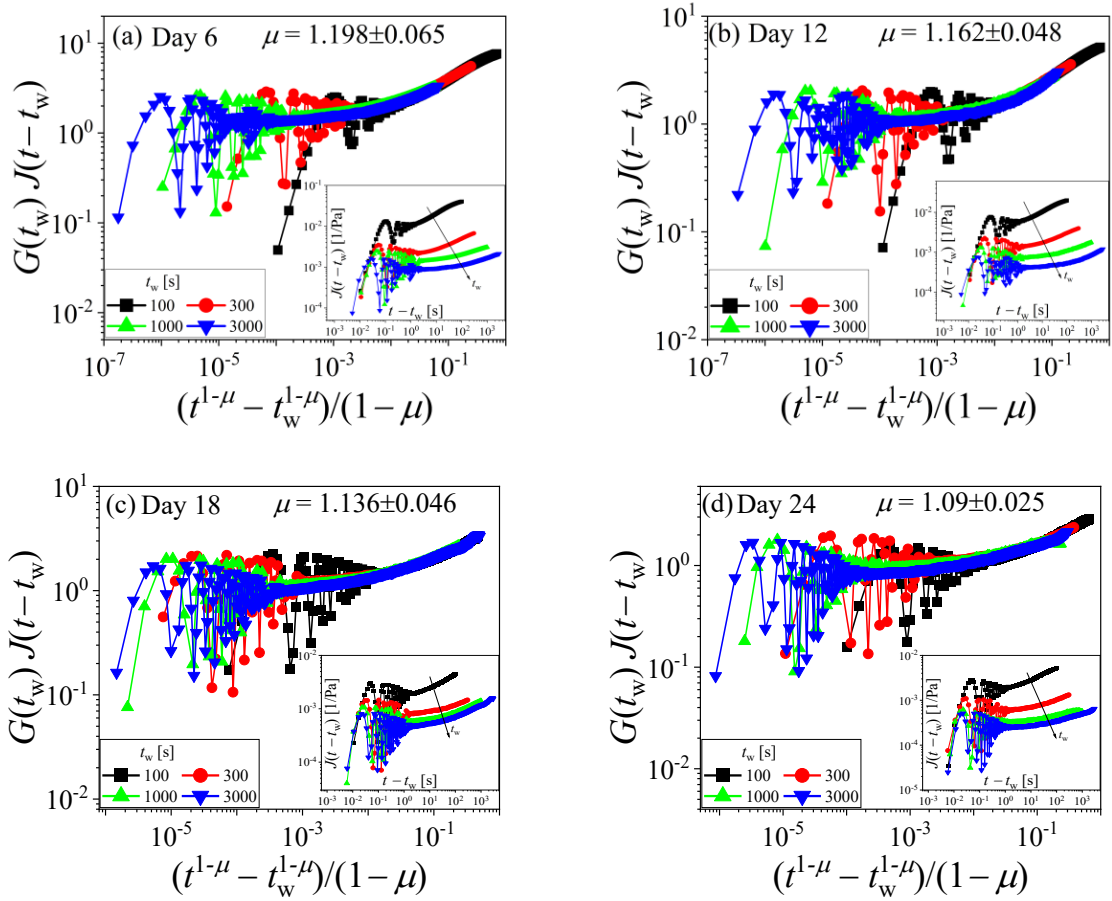

**Figure S2:** In (a – d ), the time – aging-time superposition of creep curves is plotted for $t_R =$ 6, 12, 18, and 24 days old systems, respectively. The inset plots show the creep curves obtained for different waiting times (from top to bottom: 100, 300, 1000, and 3000 s).

**Viscosity Bifurcation**

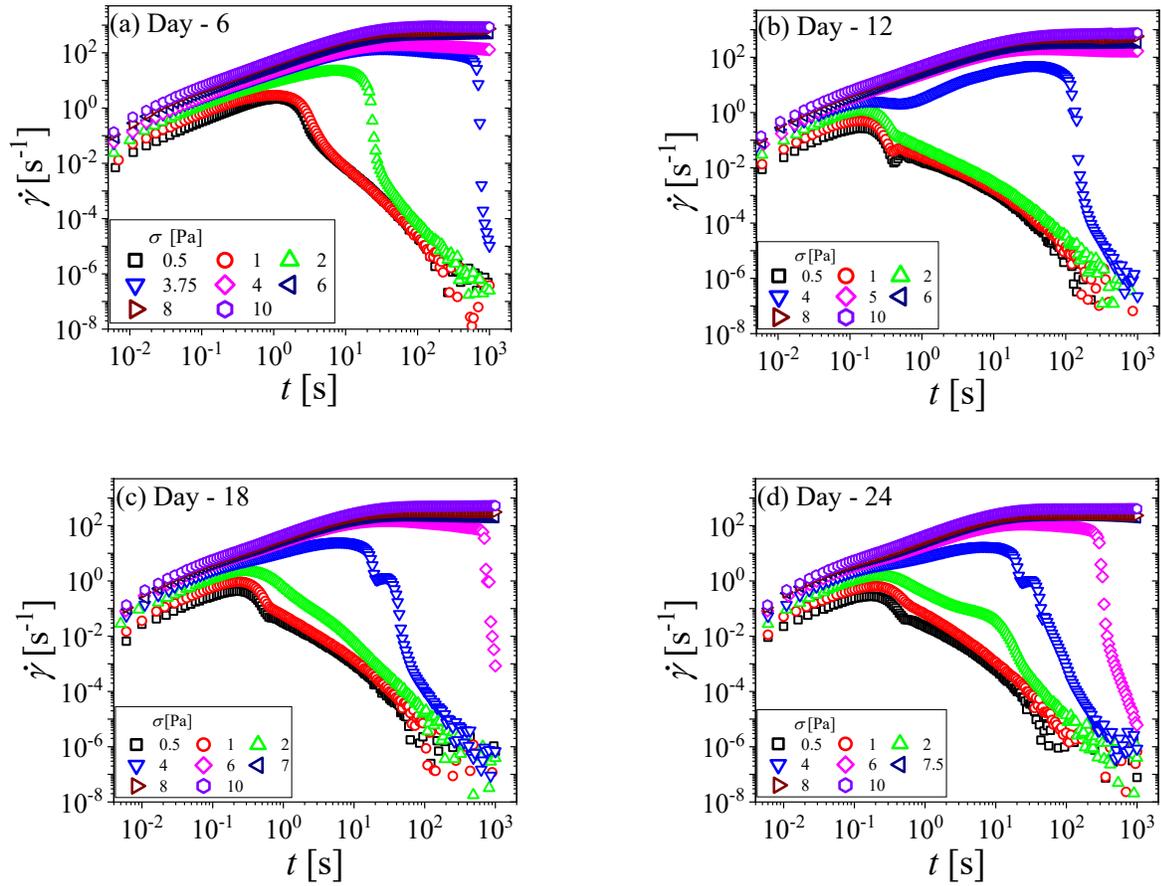

**Figure S3:** In (a – d ), the evolution of shear rate ($\dot{\gamma}$) with time for constant applied stress ($\sigma$) is plotted for $t_R$ = 6, 12, 18, and 24 days old systems. The legends show the values of different stresses applied to silica dispersion.

**Amplitude sweep**

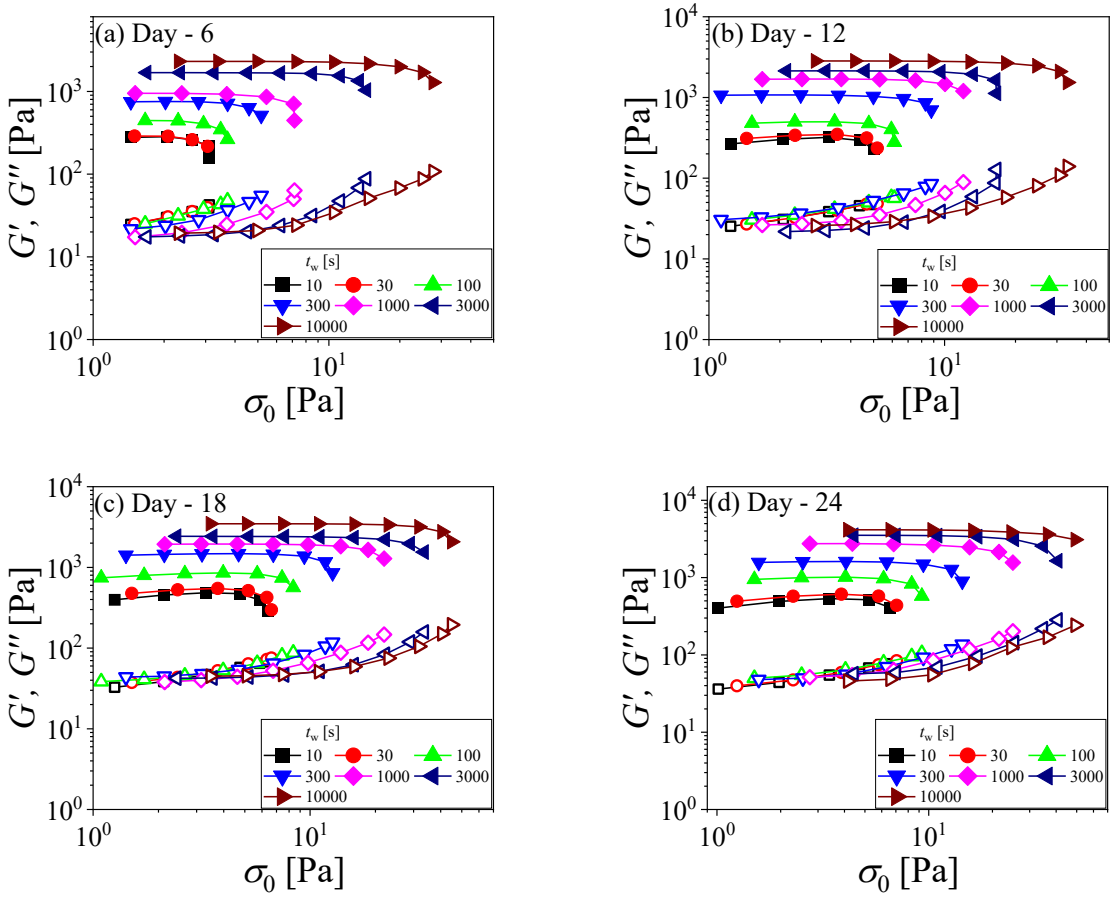

**Figure S4:** In (a – d), the $G'$ (closed symbol) and $G''$ (open symbol) are plotted against the increasing stress amplitude ($\sigma_0$) for $t_R = 6, 12, 18,$ and $24$ days old systems. The legends show the different waiting times at which amplitude sweep experiments were done.